\documentstyle[12pt,epsfig,amsfonts]{article}
\title{Self-force via energy-momentum\\ and angular momentum balance 
equations}
\author{\bf Yurij Yaremko}
\date{\it Institute for Condensed Matter Physics, \\
1 Svientsitskii St., 79011 Lviv, Ukraine}
\begin{document}
\maketitle
\begin{abstract}
The radiation reaction for a point-like charge coupled to a massive 
scalar field is considered. The retarded Green's function associated with 
the Klein-Gordon wave equation has support not only on the future light 
cone of the emission point (direct part), but extends inside the light 
cone as well (tail part). Dirac's scheme of decomposition of the retarded 
electromagnetic field into the ``mean of the advanced and retarded field'' 
and the ``radiation'' field is adapted to theories where Green's function 
consists of the direct and the tail parts. The Harish-Chandra equation of 
motion of radiating scalar charge under the influence of an external force 
is obtained. This equation includes effect of particle's own field. The 
self force produces a time-changing inertial mass.
\end{abstract}

\section{Introduction}\label{intro}
\setcounter{equation}{0}
There is an extensive literature devoted to the regularization problem in 
curved spacetime. De Witt and Brehme \cite{WB} derived the self-force 
acting upon a point electric charge in fixed background gravitational 
field. (Their expression for radiation reaction was later corrected by 
Hobbs \cite{Hb}.) The principle of equivalence in general relativity 
implies that a particle of infinitesimal mass and size moves along a 
geodesic. The particle of a small but finite mass perturbs the spacetime 
geometry. Mino, Sasaki and Tanaka \cite{MST} included the interaction of 
the particle with this perturbation which changes the worldline.
In \cite{Q} Quinn has obtained an expression for the self-force on a 
point-like particle coupled to a massless scalar field. There is a long 
series of papers devoted to the motion of a point electric charge, a point 
scalar charge, and a point mass in black hole spacetimes, where the 
effects of radiation reaction are taken into account (see references in 
review \cite{Pois}).

The computation of effect of particle's own field is not a trivial matter, 
since the Green's function associated with the wave operator has support 
within the light cone. This is because in curved spacetime massless waves 
propagate not just at a speed of light, but also at all speeds smaller 
than or equal to the speed of light. The particle may ``fill'' its own 
field, which acts on it just like an external one. The equation of motion 
require one to identify that portion of the retarded field at each point of 
the world line which arises from source contributions interior to the 
light cone. This part of field is often called the ``tail term''. The self 
force on a particle then consists of two parts: this comes from the direct 
part of the Green's function and depends on the current state of 
particle's motion and that comes from the tail part and depends not only 
the current state of the particle, but also on its past history. It leads 
to the non-local (integro-differential) equations of motion. 

Detweiler and Whiting proposed \cite{DW} a consistent decomposition of the 
retarded Green's function into singular and radiative parts. It obeys the 
spirit of Dirac's scheme of splitting of electromagnetic potential of a 
point-like charged particle arbitrarily moving in flat spacetime. Dirac 
\cite{Dir} decomposed the retarded Li\'enard-Wiechert potential $A^{\rm 
ret}$ into two parts: (i) one-half of the retarded plus one-half of the 
advanced potentials which is inhomogeneous solution of the wave equation 
$\square A_\alpha=-4\pi j_\alpha$ whose source term is infinite on the 
world line. $A^S=1/2(A^{\rm ret}+A^{\rm adv})$ is just singular as the 
retarded potential in the immediate vicinity of the particle's world line. 
The superscript ``S'' stands for ``singular'' as well as ``symmetric''. 
(ii) combination $A^R=1/2(A^{\rm ret}-A^{\rm adv})$ of one-half of the 
retarded minus one-half of the advanced potentials which satisfies the 
homogeneous wave equation. This well behaved potential can be thought as a 
free radiation field. The superscript ``R'' stands for ``radiative'' 
as well as ``regular''.

The radiative Green's function implicitly used by Dirac in flat spacetime 
is
\begin{equation}\label{Grad}
G^{\rm rad}(x,y)=G^{\rm ret}(x,y)-G^{\rm sym}(x,y)
\end{equation}
where
\begin{equation}\label{Gsym}
G^{\rm sym}(x,y)=\frac12\left[G^{\rm ret}(x,y)+G^{\rm adv}(x,y)\right].
\end{equation}

The causal structure of the Green's function is richer in curved 
spacetime. Due to contributions of the interior of the light cones, the 
retarded potential depends on the particle's history {\it prior} to the 
retarded instant $\tau^{\rm ret}(x)$ while the advanced one is generated 
by portion of particle's world line $\zeta$ {\it after} the advanced 
instant $\tau^{\rm adv}(x)$. (The retarded and the advanced moments label 
the points on $\zeta$ related with arbitrary field point $x\in{\mathbb 
M}_{\,4}$ by null rays.) The combination of half-retarded potential minus 
half-advanced one could satisfy the homogeneous wave equation. Moreover, it 
would be smooth on the world line. But a self-force constructed from this 
radiative potential will be highly non-causal. It will be depend on 
particle's entire history, both past (through the retarded Green's 
function) and future (through the advanced Green's function). The Dirac's 
scheme (\ref{Grad}) for decomposition cannot be adopted without 
modification in curved spacetime. The modification is performed in 
Ref.\cite{DW}.

Detweiler and Whiting start with a Hadamard construction of a symmetric 
scalar field Green's function
\begin{equation}\label{GsymH}
G^{\rm sym}(x,y)=\frac{1}{8\pi}\left[u(x,y)\delta(\sigma)-
v(x,y)\theta(-\sigma)\right],
\end{equation}
where $u(x,y)$ and $v(x,y)$ are smooth functions of the base point $y$ and 
the field point $x$ and $\sigma$ is half of the square of the distance 
measured along the geodesic from $x$ to $y$. The step function 
$\theta(-\sigma)$ means that this part of $G^{\rm sym}(x,y)$ has support 
within both the past null cone and the future null cone of $x$. Having 
coupled it with $\delta-$shaped distribution of scalar charge moving along a 
geodesic we obtain the tail part of the field $\psi$ as follows:
\begin{equation}
\psi_{\rm tail}^{\rm sym}(x)=-\frac{q}{2}\left[
\int_{-\infty}^{\tau^{\rm ret}(x)}{\rm d}\tau v(x,z(\tau))
+\int^{+\infty}_{\tau^{\rm adv}(x)}{\rm d}\tau v(x,z(\tau))
\right].
\end{equation}
To remove the noncausality, authors add to singular Green's function 
(\ref{GsymH}) biscalar $v(x,y)$ being solution of homogeneous wave 
equation. A new symmetric (singular) Green's function
\begin{equation}\label{GSH}
G^{\rm S}(x,y)=\frac{1}{8\pi}\left[u(x,y)\delta(\sigma)-
v(x,y)\theta(\sigma)\right],
\end{equation}
has no support within the null cone. Corresponding tail part of the scalar 
field depends on particle's history during the interval $[\tau^{\rm 
ret}(x),\tau^{\rm adv}(x)]$:
\begin{equation}\label{DWadd}
\psi_{\rm tail}^{\rm S}(x)=\frac{q}{2}\left[
\int_{\tau^{\rm ret}(x)}^{\tau^{\rm adv}(x)}{\rm d}\tau v(x,z(\tau))
\right].
\end{equation}
It is inhomogeneous solution of wave equation. Since it is singular just 
as the retarded tail field $\psi_{\rm tail}^{\rm ret}$(x), the authors 
define the radiative Green's function as follows
\begin{equation}\label{GHR}
G^{\rm R}(x,y)=G^{\rm ret}(x,y)-G^{\rm S}(x,y)
\end{equation}
(cf. eq.(\ref{Grad})). Corresponding radiation field
\begin{equation}\label{FHR}
\psi^{\rm R}(x)=-\left[\frac{qu(x,z(\tau))}{2\dot\sigma}\right]_{\tau^{\rm 
ret}(x)}^{\tau^{\rm adv}(x)}-
q\left[
\int_{-\infty}^{\tau^{\rm ret}(x)}{\rm d}\tau v(x,z(\tau))+\frac12
\int_{\tau^{\rm ret}(x)}^{\tau^{\rm adv}(x)}{\rm d}\tau v(x,z(\tau))
\right]
\end{equation}
is well behaved solution of homogeneous wave equation. In the coincidence 
limit, where field point $x$ approaches to the world line at point 
$z(\tau)$, the interval $[\tau^{\rm ret}(x),\tau^{\rm adv}(x)]$ shrinks to 
zero. Due to taking of this limit the potential (\ref{FHR}) is generated by 
the portion of the world line that corresponds to the causal interval 
$]-\infty,\tau]$. The last integral in eq. (\ref{FHR}) gives no contribution 
to a self-force because it cancels the ill defined part of gradient 
$\triangle_\alpha\psi_{\rm tail}^{\rm ret}$ coming from implicit dependence 
of $\tau^{\rm ret}$ upon $x$.

Following this scheme, Detweiler and Whiting recovered the results 
\cite{WB,MST,Q} for electromagnetic, scalar, and gravitational fields.

It is obvious that the physically relevant solution of the wave equation 
is the retarded potential. Teitelboim \cite{Teit} derived the 
electromagnetic self-force in flat spacetime within the framework of 
retarded causality. The author substituted the retarded Li\'enard-Wiechert 
field in the Maxwell energy-momentum tensor density and calculated the 
flow of energy-momentum which flows across a space-like surface. Minkowski 
space was parameterized by four curvilinear coordinates. The first, proper 
time, labels points of emission placed on $\zeta$, the second one 
determines the surface (e.g., a tilted hyperplane which is orthogonal to 
particle's 4-velocity at fixed instant of observation). Having integrated 
the stress-energy tensor over two angular variables that distinguish 
points on the surface, Teitelboim found the flow of energy-momentum 
mentioned above. The resulted expression depends on the particle's 
individual characteristics (on its mass, its charge, its velocity and 
acceleration). In fact, the surface integration is equivalent to taking of 
coincidence limit in Dirac's scheme. Abraham-Lorentz-Dirac expression for 
electromagnetic self-force is obtained in \cite{Teit} via consideration of 
energy-momentum conservation. 
 
Teitelboim \cite{Teit} demonstrates that each of two terms which constitute 
Abraham radiation-reaction four-vector originates from the specific part of 
Maxwell energy-momentum tensor density. The author decomposes the 
stress-energy tensor into two divergent-free components: $\hat T={\hat 
T}_{\rm bnd}+{\hat T}_{\rm rad}$. Surface integration of the bound part 
${\hat T}_{\rm bnd}$ results the flow which never gets far from the 
point-like source. It is permanently ``attached'' to the charge and is 
carried along with it. A charged particle cannot be separated from its bound 
electromagnetic ``cloud'' which has its own 4-momentum and angular momentum. 
``Bare'' charge and electromagnetic ``cloud'' constitute new entity: dressed 
charged particle. ${\hat T}_{\rm bnd}$ contributes into particle's inertia: 
4-momentum of dressed charge $e$ contains, apart from usual velocity term, 
the extra term whose time derivative is exactly the negative of the Schott 
term:
\begin{equation}\label{p_a}
p^\mu = mu^\mu -\frac{2e^2}{3}a^\mu.
\end{equation}
Calculation of flux of the radiative component ${\hat T}_{\rm rad}$ yields 
the Larmor relativistic rate of radiated energy-momentum. This part of 
energy-momentum detaches itself from the charge and leads an independent 
existence. 

In this paper we tend Teitelboim's approach to allow contributions from 
interior of the light cone. For the clearest demonstration of the impact 
of our analysis, we refer to a point-like particle of mass $m$ and charge 
$e$ coupled to electromagnetic field in flat spacetime of three 
dimensions \cite{Yar3D,Y3D}. In $2+1$ electrodynamics the retarded Green's 
function associated with D'Alembert operator 
$\square=\eta^{\alpha\beta}\partial_\alpha\partial_\beta$ is supported 
within the light cone \cite{Gl,KLS}:
\begin{equation}\label{G3ret}
G^{\rm ret}_{2+1}(x,y)=\frac{\theta(x^0-y^0-|{\mathbf x}-{\mathbf 
y}|)}{\sqrt{-2\sigma(x,y)}}.
\end{equation}
$\theta(x^0-y^0-|{\mathbf x}-{\mathbf y}|)$ is the light cone step 
function defined to be one if $x^0-y^0\ge|{\mathbf x}-{\mathbf y}|$ and 
defined to be zero otherwise. Synge's world function in flat space-time is 
numerically equal to half the squared distance between $x$ and $y$:
\cite{Pois}
\begin{equation}\label{Sng}
\sigma(x,y)=\frac12\eta_{\alpha\beta}(x^\alpha -y^\alpha)(x^\beta 
-y^\beta).
\end{equation}
The analysis of the simplest model with tails will give more deep 
understanding of Detweiler and Whiting scheme of decomposition. 

The paper is organized as follows. In Section \ref{PF} and \ref{field-sc}, 
we define the scalar potentials and scalar field strengths. In Section 
\ref{Sc-rad} we split the energy-momentum and angular momentum carried 
by massive scalar field into bound and radiative parts. Extracting of 
radiated portions of Noether quantities is not a trivial matter, since the 
massive field holds energy and momentum near the source. In Section 
\ref{meq}, we derive equation of motion of radiating scalar pole via 
analysis of  energy-momentum and angular momentum balance equations. It is 
of great importance that conservation laws yield the Harish-Chandra 
equation of motion \cite{HC}. In Section \ref{concl}, we discuss the result 
and its implications.

\section{Green's function and potentials}\label{PF}
\setcounter{equation}{0}
The dynamics of a point-like charge coupled to massive scalar field is 
governed by the action \cite{Kos,Pois}
\begin{equation}\label{Itot}
I_{\rm total}=I_{\rm part}+I_{\rm int}+I_{\rm field}.
\end{equation}
Here
\begin{equation}
I_{\rm field}=-\frac{1}{8\pi}\int{\rm d}^4x\left(
\eta^{\alpha\beta}\varphi_\alpha\varphi_\beta+k_0^2\varphi^2\right)
\end{equation}
is an action functional for a massive scalar field $\varphi$ in flat 
spacetime with metric tensor $\eta^{\alpha\beta}={\rm diag}(-1,1,1,1)$. The 
mass parameter $k_0$ is a constant with the dimension of reciprocal 
length. Its physical sense will be discussed below. The integration is 
performed over all the spacetime. The particle action is
\begin{equation}\label{SS}
I_{\rm part}=-m_0\int{\rm d}\tau\sqrt{-{\dot z}^2}
\end{equation}
where $m_0$ is the bare mass of the particle which moves on a world line 
$\zeta\in{\mathbb M}_{\,4}$ described by relations $z^\alpha(\tau)$ which 
give the particle's coordinates as functions of proper time; ${\dot 
z}^\alpha(\tau)={\rm d} z^\alpha(\tau)/{\rm d}\tau$. Finally, the 
interaction term is given by
\begin{equation}
I_{\rm int}=g\int{\rm d}\tau\sqrt{-{\dot z}^2}\varphi(z)
\end{equation}
where $g$ is scalar charge carried by a point-like particle. 

Variation on field variable $\varphi$ of action (\ref{Itot}) yields 
the Klein-Gordon wave equation
\begin{equation}\label{KG}
\left(\square -k_0^2\right)\varphi(x)=-4\pi\rho (x),
\end{equation}
where $\square=\eta^{\alpha\beta}\partial_\alpha\partial_\beta$ is the 
D'Alembert operator. We consider a scalar field $\varphi(x)$ with a point 
particle source
\begin{equation}\label{msrd}
\rho(x)=g\int_{-\infty}^{+\infty}{\rm d}\tau \delta^4(x-z(\tau)),
\end{equation}
where $g$ is coupling constant and $\delta^4(x-z(\tau))$ is a 
four-dimensional Dirac's distribution localized on the world line: 
charge's density is zero everywhere, except at the particle's position 
where it is infinite. 

The relevant wave equation for the Green's function $G(x,y)$ is
\begin{equation}\label{Gwe}
\left(\square -k_0^2\right)G(x-y)=-\delta^4(x-y).
\end{equation}
Its solution is the symmetric Green's function \cite{Pois,Kos}
\begin{equation}\label{GmS}
G^{\rm sym}(x-y)=\frac{1}{4\pi}\left[
\delta(\sigma)-\frac{k_0}{\sqrt{-2\sigma}}J_1(k_0\sqrt{-2\sigma})
\theta(-\sigma)\right],
\end{equation}
where $J_1$ is the first order Bessel's function whose argument contains 
Synge's world function (\ref{Sng}).

\begin{figure}[ht]
\begin{center}
\epsfclipon
\epsfig{file=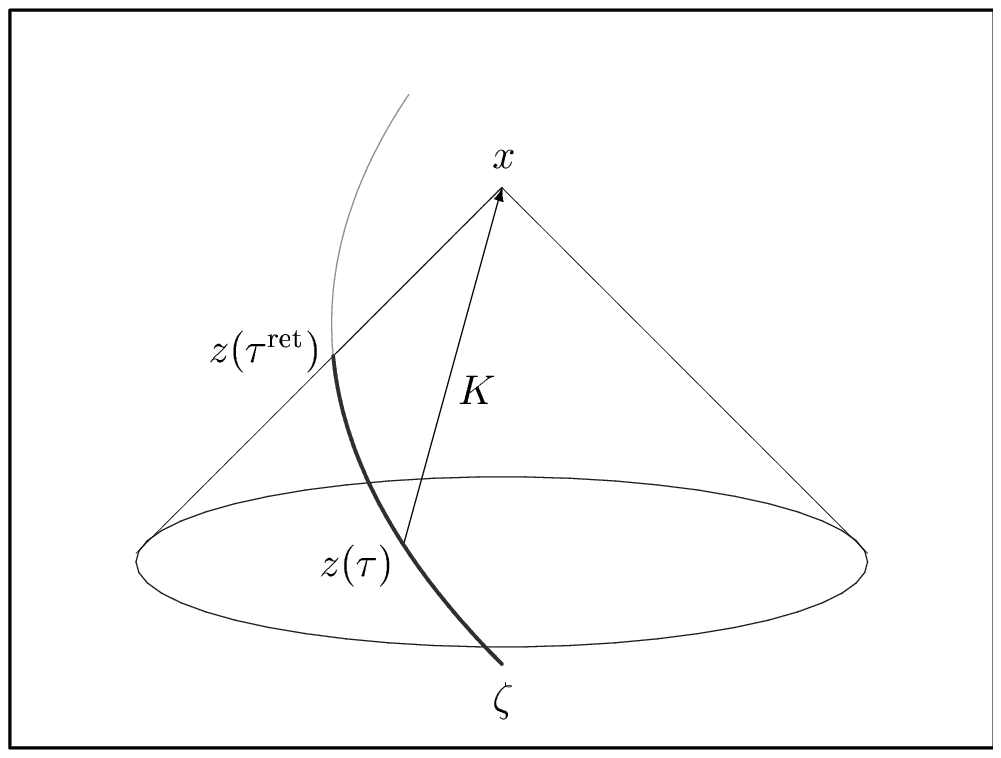,width=0.43\textwidth}
\end{center}
\caption{Direct part of the retarded potential at field point $x$ is 
generated by the intersection $z(\tau^{\rm ret})$ of the world line $\zeta$ 
and past light cone with vertex at $x$. Tail part of $\varphi^{\rm ret}(x)$ 
depends on the segment of $\zeta$ that corresponds to time interval 
$]-\infty,\tau^{\rm ret}(x)]$. The tail term defines contribution from 
cone's interior. The vector $K$ is a vector pointing from the point of 
emission $z(\tau)\in\zeta$ to a field point $x$.
}\label{ret1}
\end{figure}

Convoluting the retarded Green's function 
\begin{eqnarray}\label{GmscR}
G^{\rm ret}(x-y)&=&\theta(x^0-y^0)G^{\rm sym}(x-y)\\
&=&\frac{1}{4\pi}\left[
\frac{\delta(T-R)}{R}-
\theta(T-R)\frac{k_0J_1(k_0\sqrt{T^2-R^2})}{\sqrt{T^2-R^2}}
\right],\nonumber
\end{eqnarray}
where $T=x^0-y^0$ and $R=|{\mathbf x}-{\mathbf y}|$ with charge density 
(\ref{msrd}), we construct the massive scalar field. The delta-function in 
eq.(\ref{GmscR}) results direct term which is generated by a single event in 
space-time: the intersection of the world line and $x$'s past light cone. 
The Heaviside stepfunction extends the support to the segment of $\zeta$ 
that corresponds to the interval $\tau\in]-\infty,\tau^{\rm ret}(x)]$. 
Summing up the direct term and the tail term, we construct the retarded 
potential produced by an arbitrarily moving point-like source coupled to 
massive scalar field \cite{Kos,HC,H1}:
\begin{equation}\label{sg-sm}
\varphi^{\rm ret}(x)=\frac{g}{r}-g\int\limits_{-\infty}^{\tau^{\rm 
ret}(x)} {\rm d}\tau \frac{k_0J_1[k_0\sqrt{-(K\cdot K)}]}{\sqrt{-(K\cdot 
K)}}.
\end{equation}
The first term is referred to the retarded point $z^{\rm ret}$ associated 
with $x$; $r$ is the retarded distance between $z^{\rm ret}$ and $x$.  
$K=x-z(\tau)$ is four-vector pointing from $z(\tau)\in\zeta$ to a point 
$x$ where the potential $\varphi$ is observed (see Figure \ref{ret1}). 

\begin{figure}[ht]
\begin{center}
\epsfclipon
\epsfig{file=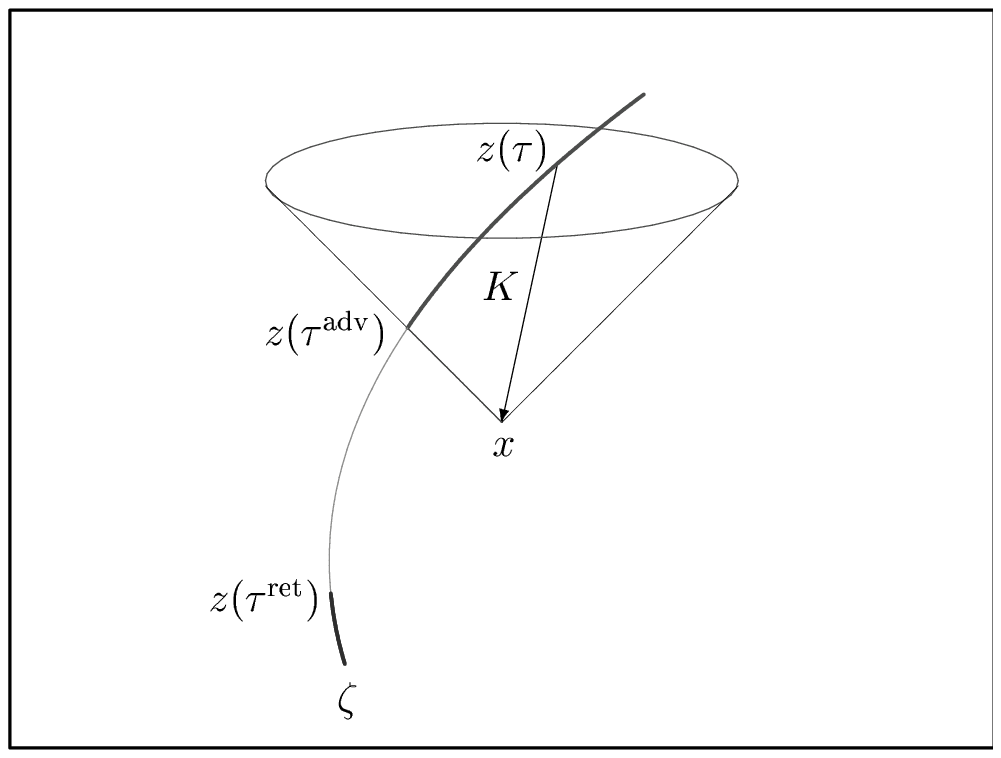,width=0.4\textwidth}
\end{center}
\caption{The advanced Green's function (\ref{GmscA}) possesses support on 
the surface of the future light cone of $x$ (direct part) as well as within 
this cone (tail part). The portion of the world line after $\tau^{\rm 
adv}(x)$ produces the advanced potential (\ref{sg-adv}). Advanced instant 
labels point at which particle's world line punctures future light cone 
with vertex at $x$.
}\label{adv1}
\end{figure}

Convolution of the advanced Green's function 
\begin{eqnarray}\label{GmscA}
G^{\rm adv}(x-y)&=&\theta(-x^0+y^0)G^{\rm sym}(x-y)\\
&=&\frac{1}{4\pi}\left[
\frac{\delta(T+R)}{R}-
\theta(-T-R)\frac{k_0J_1(k_0\sqrt{T^2-R^2})}{\sqrt{T^2-R^2}}
\right]\nonumber
\end{eqnarray}
coupled with the $\delta$-like density (\ref{msrd}) is analogous to the 
manipulations with its retarded counterpart. The advanced potential
\begin{equation}\label{sg-adv}
\varphi^{\rm adv}(x)=\frac{g}{r_
{\rm adv}}-g\int\limits^{+\infty}_{\tau^{\rm 
adv}(x)} {\rm d}\tau
\frac{k_0J_1[k_0\sqrt{-(K\cdot K)}]}{\sqrt{-(K\cdot K)}}
\end{equation}
is generated by the point charge during its entire future history following 
the advanced time associated with $x$ (see Figure \ref{adv1}). Particle's 
characteristics in the direct term are referred to the instant $\tau^{\rm 
adv}(x)$.

The simplest scalar field is generated by an unmoved source placed at 
the coordinate origin \cite{Bha,CM}:
\begin{equation}\label{YuP}
\varphi_{\rm Yu}(x)=g\frac{\exp(-k_0r)}{r}.
\end{equation}
Here $r=\sqrt{(x^1)^2+(x^2)^2+(x^3)^2}$ is the distance to the charge. 
It is the well-known Yukawa potential which plays an important role in the 
theory of nucleon-nucleon interactions. 

\section{Scalar field strengths}\label{field-sc}
\setcounter{equation}{0}

Let two particles interact through a massive scalar field. Static charge 
$g_1$ placed at the coordinate origin generates the Yukawa field with 
components
\begin{eqnarray}
\Phi_i(x)&=&\frac{\partial\varphi_{\rm Yu}}{\partial x^i}\\
&=&-g_1\frac{e^{-k_0r}}{r^2}\left(1+k_0r\right)n_i,\nonumber
\end{eqnarray}
where $n_i=x_i/r$ is unit direction vector and $r$ is the distance to the 
charge. This charge exerts another one, say $g_2$, with the force
\begin{equation}\label{YukFrc}
{\mathbf 
F}_{12}=-g_1g_2\frac{e^{-k_0r_{12}}}{r_{12}^2}\left(1+k_0r\right)
{\mathbf n}_{12},
\end{equation}
where ${\mathbf n}_{12}$ is unit vector codirectional with the radius 
vector ${\mathbf r}_{12}$ drawing from particle $1$ to particle $2$. The 
force is attractive if $g_1$ and $g_2$ are of like sign. The charges can be 
interpreted as nucleons which attract each other and are bound in a 
nucleus. The force (\ref{YukFrc}) elucidates also why a neutron repels an 
antineutron. 

In general, the action (\ref{Itot}) can be considered as a classical 
version of the Yukawa model for strongly interacting nucleons. It is 
assumed that the neutrons and protons are joint together by a massive 
pseudoscalar pion field. Parameter $k_0$ associates with the rest mass of a 
massive scalar field particle mediating the interaction. The parameter acts 
as a cutoff: the heavier is the field particle, the shorter is range of 
Yukawa force. Yukawa force is essential at a distance about $1/k_0$.

Let us derive the field produced by an arbitrarily moving scalar charge.
Scalar field strengths are given by the gradient of the potential 
(\ref{sg-sm}). Differentiation of the direct term yields
\begin{equation}\label{Psrt}
\frac{\partial\varphi^{\rm ret}_{\rm dir}(x)}{\partial 
x^\mu}=-g\left[
\frac{[1+(K\cdot a)]K_\mu}{r^3}-\frac{u_\mu}{r^2}
\right]_{\tau=\tau^{\rm ret}(x)}.
\end{equation}
Further we differentiate the tail term in the potential (\ref{sg-sm}) which 
arises from source contributions interior to the light cone. Apart from the 
integral
\begin{equation}\label{fth}
\Phi_\mu^{(\theta)}(x)=g\int\limits_{-\infty}^{\tau^{\rm ret}(x)} {\rm 
d}\tau k_0^2
\frac{{\rm d}}{{\rm d}W}\left(\frac{J_1(W)}{W}\right)k_0
\frac{K_\mu}{\sqrt{-(K\cdot K)}}
\end{equation}
the gradient $\Phi_{{\rm tail},\mu}=\Phi_\mu^{(\theta)}+\Phi_\mu^{(\delta)}$ 
contains also local term
\begin{equation}\label{fdl}
\Phi_\mu^{(\delta)}(x)=gk_0^2\left.\frac{J_1(W)}{W}
\frac{K_\mu}{r}\right|_{\tau=\tau^{\rm ret}}
\end{equation}
which is due to time-dependent upper limit of integral in eq.(\ref{sg-sm}). 
Argument of Bessel's function $W:=k_0\sqrt{-(K\cdot K)}$.

To simplify the tail contribution us much us possible we use the identity
\begin{equation}
\frac{k_0}{\sqrt{-(K\cdot K)}}=\frac{1}{(K\cdot u)}
\frac{{\rm d}W}{{\rm d}\tau}
\end{equation}
in the integral (\ref{fth}) and perform integration by parts. On 
rearrangement, we add it to the expression (\ref{fdl}). The term which 
depends on the end points only annuls $f_\mu^{(\delta)}$. Finally, the 
gradient of tail term of the potential (\ref{sg-sm}) becomes
\begin{equation}\label{grad-tl}
\frac{\partial\varphi^{\rm ret}_{\rm tail}(x)}{\partial 
x^\mu}=
g\int\limits_{-\infty}^{\tau^{\rm ret}(x)} {\rm d}\tau 
k_0^2\frac{J_1(W)}{W}\left[\frac{1+(K\cdot a)}{r^2}K_\mu-
\frac{u_\mu}{r}\right].
\end{equation}
The particle's position, velocity, and acceleration under the integral sign 
are evaluated at instant $\tau\le \tau^{\rm ret}(x)$. The invariant quantity
\begin{equation}\label{rint-sc}
r=-(K\cdot u)
\end{equation}
is an affine parameter on the time-like geodesic that links $x$ to 
$z(\tau)$; it can be loosely interpreted as the time delay between $x$ and
$z(\tau)$ as measured by an observer moving with the particle. (For such 
hypothetic observer the particle is momentarily at rest: its four-velocity
$u^\alpha=(1,0,0,0)$.)

Because of asymptotic behavior of the first order Bessel's function with 
argument $W:=k_0\sqrt{-(K\cdot K)}$ the retarded field
\begin{eqnarray}\label{Fmret}
\Phi_\mu^{\rm ret}(x)&=&
\frac{\partial\varphi^{\rm ret}(x)}{\partial x^\mu}\\
&=&-g\frac{1+(K\cdot a)}{r^3}K_\mu+g\frac{u_\mu}{r^2} + 
g\int\limits_{-\infty}^{\tau^{\rm ret}(x)} {\rm d}\tau 
k_0^2\frac{J_1(W)}{W}\left[\frac{1+(K\cdot a)}{r^2}K_\mu-
\frac{u_\mu}{r}\right]\nonumber
\end{eqnarray}
is finite on the light cone where $W=0$. It diverges on the particle's 
trajectory only. Indeed, if the point of emission $z(\tau)$ approaches the 
field point $x$, all the components of four-vector $K=x-z(\tau)$ becomes 
infinitesimal ones and the distance (\ref{rint-sc}) tends to zero.

The advanced field
\begin{eqnarray}\label{Fmadv}
\Phi_\mu^{\rm adv}(x)&=&
\frac{\partial\varphi^{\rm adv}(x)}{\partial x^\mu}\\
&=&-g\frac{1+(K\cdot 
a)}{(r_{\rm adv})^3}K_\mu-g\frac{u_\mu}{(r_{\rm adv})^2} + 
g\int\limits^{+\infty}_{\tau^{\rm adv}(x)} {\rm d}\tau 
k_0^2\frac{J_1(W)}{W}\left[\frac{1+(K\cdot a)}{r^2}K_\mu-
\frac{u_\mu}{r}\right]\nonumber
\end{eqnarray}
depends on particle's future history (see Figure \ref{adv1}). Particle's 
characteristics in the direct part are referred to the advanced instant 
$\tau^{\rm adv}(x)$.

\section{Scalar radiation}\label{Sc-rad}
\setcounter{equation}{0}
The action (\ref{Itot}) is invariant under infinitesimal transformations 
(translations and rotations) which constitute the Poincar\'e group. 
According to Noether's theorem, these symmetry properties yield 
conservation laws, i.e. those quantities that do not change with time. 
Outgoing radiation removes energy, momentum, and angular momentum from the 
source. The quantities are defined in standard way \cite{Rohr}, as flows of 
the stress-energy tensor density $\hat T$
\begin{equation}\label{sc-p}
p_{\rm sc}^\nu(\tau)=\int_{\Sigma}{\rm d}\sigma_\mu T_{\rm sc}^{\mu\nu},
\end{equation}
and its torque
\begin{equation}\label{sc-M}
M_{\rm sc}^{\mu\nu}(\tau)=\int_{\Sigma}{\rm d}\sigma_\alpha\left(x^\mu 
T_{\rm sc}^{\nu\alpha}-x^\nu T_{\rm sc}^{\mu\alpha}\right),
\end{equation}
that flow across a space-like surface $\Sigma$ which intersects a world line
at point $z(\tau)$; ${\rm d}\sigma_\mu$ is the vectorial surface element on 
$\Sigma$. While the stress-energy tensor is quite different than that in 
classical electrodynamics:
\begin{equation}\label{T}
4\pi T_{\mu\nu}=\frac{\partial\varphi}{\partial x^\mu}
\frac{\partial\varphi}{\partial x^\nu}-\frac{\eta_{\mu\nu}}{2}
\left(
\eta^{\alpha\beta}\frac{\partial\varphi}{\partial x^\alpha}
\frac{\partial\varphi}{\partial x^\beta}+k_0^2\varphi^2
\right),
\end{equation}
(see Refs.\cite{CM,Cw69,Cw70}). To prepare the way for our discussion of 
a self-force acted on a point-like source coupled with massive scalar 
field, we consider first the relatively simple case of a massless scalar 
field.

\subsection{Massless scalar radiation}\label{massless}

Let parameter $k_0=0$. The rate of change of the energy-momentum of the 
retarded field is computed by means of the stress-energy tensor (\ref{T}) 
where field tensor components are replaced by the direct field strengths 
(\ref{Psrt}). We call the part of massless scalar field which scales as 
$r^{-1}$ the radiative part of field: \begin{equation}\label{frad}
\Phi_\mu^{\rm rad}(x)=-\frac{e}{r}(a\cdot k)k_\mu.
\end{equation}
(Null vector $k^\mu=K^\mu/r$ is normalized in such a way that scalar product 
$(k\cdot u)=-1$.) The part which scales as $r^{-2}$ does not involve the 
acceleration:
\begin{equation}\label{fbnd}
\Phi_\mu^{\rm bnd}(x)=-\frac{e}{r^2}\left(k_\mu-u_\mu\right).
\end{equation} 
This is the bound part of field. Following Teitelboim \cite{Teit}, we 
decompose the stress-energy tensor into bound and radiative parts which are 
separately conserved off the world line. To calculate the radiative part it 
is straightforward to substitute eq.(\ref{frad}) into eq.(\ref{T}):
\begin{equation}\label{sc-Trad}
4\pi T_{\rm rad}^{\mu\nu}=\frac{g^2}{r^2}(a\cdot k)^2k^\mu k^\nu.
\end{equation}
The bound part
\begin{equation}\label{sc-Tbnd}
4\pi T_{\rm bnd}^{\mu\nu}=T_{(-4)}^{\mu\nu}+T_{(-3)}^{\mu\nu}
\end{equation}
is the combinations of terms which depend on the retarded distance as 
$r^{-4}$ and $r^{-3}$, respectively:
\begin{eqnarray}\label{sf-T4}
4\pi T_{(-4)}^{\mu\nu}&=&\frac{g^2}{r^4}\left(
k^\mu k^\nu-k^\mu u^\nu-u^\mu k^\nu+u^\mu u^\nu-\frac12\eta^{\mu\nu}
\right),\\
4\pi T_{(-3)}^{\mu\nu}&=&\frac{g^2}{r^3}(a\cdot 
k)\left(\phantom{\frac11}\!\!\!\!
2k^\mu k^\nu-k^\mu u^\nu-u^\mu k^\nu-\eta^{\mu\nu}\right).\label{sf-T3}
\end{eqnarray}
To decompose the angular momentum tensor density 
\begin{equation}\label{M}
M^{\mu\nu\alpha}=x^\mu T^{\nu\alpha}-x^\nu T^{\mu\alpha}.
\end{equation} 
into the bound component and the radiative component, we use the formulae 
presented firstly in pioneer work \cite{LV}:
\begin{eqnarray}\label{Mbnd}
M_{\rm bnd}^{\mu\nu\alpha}&=&z^\mu T_{\rm bnd}^{\nu\alpha} - 
z^\nu T_{\rm bnd}^{\mu\alpha} +
K^\mu T_{\rm (-4)}^{\nu\alpha} - K^\nu T_{\rm (-4)}^{\mu\alpha}, 
\\ 
M_{\rm rad}^{\mu\nu\alpha}&=&x^\mu T_{\rm rad}^{\nu\alpha} - 
x^\nu T_{\rm rad}^{\mu\alpha} +
K^\mu T_{\rm (-3)}^{\nu\alpha} - K^\nu T_{\rm (-3)}^{\mu\alpha}. 
\label{Mrad}
\end{eqnarray}

We enclose particle's world line by a very thin tube \cite{Bha} of constant 
radius $r$ and calculate fluxes of the stress-energy tensor (\ref{sc-p}) and 
its torque (\ref{sc-M}) through this surface. An appropriate coordinates are 
the retarded coordinates \cite[Chapt.II]{Pois} locally given by
\begin{equation}\label{coord}
x^\alpha=z^\alpha(s)+rk^\alpha.
\end{equation}
Here $z(s)\in\zeta$ is the retarded point associated with a field point 
$x$, $r$ is the retarded distance between these points, and $k^\alpha(x)$ 
is a null vector field tangent to the congruence of null rays that emanate 
from $z(s)$.

Globally, the flat space-time ${\Bbb M}_{\,4}$ can be thought as a disjoint 
union of world tubes of all possible radii $r>0$. A world tube is a 
disjoint union of spheres of constant radii $r$ centered on a world line of 
the particle. The sphere $S(z(s),r)$ is the intersection of the future 
light cone generated by null rays emanating from point
$z(s)\in\zeta$ in all possible directions
\begin{equation}\label{C0}
C^+(z(s))=\{x\in {\Bbb M}_{\,4}: 
\eta_{\alpha\beta}\left(x^\alpha-z^\alpha(s)\right)
\left(x^\beta-z^\beta(s)\right)=0, x^0-z^0(s)>0\}
\end{equation}
and the tilted hyperplane
\begin{equation}\label{Ksi}
\Sigma(z(s),r)=\{x\in {\Bbb M}_{\,4}: 
u_\alpha(s)\left(x^\alpha 
-z^\alpha(s)-u^\alpha(s)r\right)=0\}.
\end{equation}
Points on the sphere $S(z(s),r)=C^+(z(s))\cap\Sigma(z(s),r)$ are 
distinguished by pair of angles: polar angle $\vartheta$ and azimuthal one 
$\varphi$, which specify direction of null vector $k$ 
(see. \cite[Chapt.II, Fig.7]{Pois}).

A point in Minkowski space can be specified therefore by means of 
curvilinear coordinates $(s,r,\vartheta,\varphi)$. To define a retarded 
coordinate system, we choose an origin point $z(s)$ on particle's world line,
future light cone (\ref{C0}) with vertex at this point, and tilted 
hyperplane (\ref{Ksi}) orthogonal to particle's four-velocity $u(s)$. The 
relation between Minkowski coordinates and retarded coordinates is given by 
eq.(\ref{coord}).

We restrict ourselves to calculation of radiative parts of energy, 
momentum, and angular momentum. The technique developed in Ref.\cite{PoisPr}
can be easily adapted to this task. The outward directed surface element 
${\rm d}\sigma_\mu$ of the cylinder $r=const$ is
\begin{equation}\label{dsgm}
{\rm d}\sigma_\mu=r^2\left[-u_\mu+(1+ra_k)k_\mu\right]{\rm d}s\wedge{\rm 
d}\Omega,
\end{equation}
where $a_k:=(a\cdot k)$ is the component of particle's acceleration in the 
direction $k$ and ${\rm d}\Omega=\sin\vartheta{\rm d}\vartheta\wedge{\rm 
d}\varphi$ is the element of a solid angle. The radiative part of 
energy-momentum carried by the massless scalar field is
\begin{eqnarray}\label{p_sc}
p_{\rm dir,R}^\nu(\tau)&=&\int_{\Sigma_r}{\rm d}\sigma_\mu T_{\rm 
rad}^{\mu\nu}\\
&=&\frac{1}{4\pi}\int\limits_{-\infty}^\tau{\rm d}s
\int{\rm d}\Omega r^2\left[-u_\mu+(1+ra_k)k_\mu\right]
\frac{g^2}{r^2}(a\cdot k)^2k^\mu k^\nu\nonumber\\
&=&\frac{g^2}{3}\int\limits_{-\infty}^\tau{\rm d}sa^2(s)u^\nu(s).\nonumber
\end{eqnarray}
Integration over angular variables is handled via the relations 
\cite{PoisPr}:
\begin{eqnarray}\label{k-Om}
&&\frac{1}{4\pi}\int{\rm d}\Omega=1,\\
&&\frac{1}{4\pi}\int{\rm d}\Omega k^\alpha=u^\alpha,\nonumber\\
&&\frac{1}{4\pi}\int{\rm d}\Omega 
k^\alpha k^\beta=\frac{4}{3}u^\alpha u^\beta+
\frac13\eta^{\alpha\beta},\nonumber\\
&&\frac{1}{4\pi}\int{\rm d}\Omega k^\alpha k^\beta k^\gamma=
2u^\alpha u^\beta u^\gamma +\frac13 u^\alpha\eta^{\beta\gamma} 
+\frac13 u^\beta\eta^{\alpha\gamma}+\frac13 u^\gamma\eta^{\alpha\beta}
.\nonumber
\end{eqnarray} 

The computation of radiated angular momentum is virtually 
identical to that presented above and we do not bother with details. 
Resulting expression is as follows:
\begin{eqnarray}\label{M_sc}
M_{\rm dir,R}^{\mu\nu}(\tau)&=&\frac{g^2}{3}\int_{-\infty}^\tau{\rm d} 
s\left\{
\phantom{\frac11}\!\!\!\!
a^2(s)\left[\phantom{\frac11}\!\!\!\!
z^\mu(s)u^\nu(s)-z^\nu(s)u^\mu(s)\right]\right.\\
&+&\left.u^\mu(s)a^\nu(s)-u^\nu(s)a^\mu(s)\phantom{\frac11}\!\!\!\!\right\}.
\nonumber
\end{eqnarray}
We see that the radiated energy-momentum carried by massless scalar field 
is equal to one-half of the well-known Larmor rate of radiation integrated 
over the world line. Similarly, the radiated angular momentum is equal to 
the one-half of corresponding quantity in classical electrodynamics 
\cite{LV}. This finding is in line with expressions obtained in 
\cite{HC,H1,BV} for massless scalar self-force.

\subsection{Radiation carried by massive scalar field}\label{massive}

In this Section we decompose the energy-momentum and angular momentum 
carried by massive scalar field into the bound and radiative parts. The 
bound terms will be absorbed by particle's individual characteristics 
while the radiative terms exert the radiation reaction. We do not calculate 
the flows of the massive scalar field across a thin tube around a world line 
of the source. To extract the appropriate finite parts of energy-momentum 
and angular momentum we apply the scheme developed in Refs.\cite{Yar3D,Y3D}. 
The scheme summarizes cumbersome calculations of flows of energy, momentum, 
and angular momentum carried by electromagnetic field of a point-like 
source arbitrarily moving in flat space-time of three dimensions. In 
$2+1$ electrodynamics both the electromagnetic potential and the
electromagnetic field are non-local: they depend not only on the current 
state of motion of the particle, but also on its past (or future) history. 
The scalar field strengths (\ref{Fmret}) and (\ref{Fmadv}) behave 
analogously.

To find the tail parts of radiated Noether quantities sourced by the 
interior of light cone we deal with the retarded and the advanced fields 
defined on the world line only:
\begin{eqnarray}\label{tlFret}
F_\mu^{\rm ret}&=&g\int_{-\infty}^{\tau_1}{\rm d}\tau_2 f(z_1,z_2),\\
F_\mu^{\rm adv}&=&g\int^{\tau}_{\tau_1}{\rm d}\tau_2 f(z_1,z_2).
\label{tlFadv}
\end{eqnarray}
The integrand is 
\begin{equation}\label{f_m}
f_\mu(z_1,z_2)=gk_0\frac{J_1[k_0\sqrt{-(q\cdot q)}]}{\sqrt{-(q\cdot q)}}
\left[\frac{1+(q\cdot a_2)}{r_2^2}q_\mu- \frac{u_{2,\mu}}{r_2}\right]
\end{equation}
where $w=k_0\sqrt{-(q\cdot q)}$. Time-like vector $q^\mu=z_1^\mu-z_2^\mu$ 
connects an emission point $z(\tau_2)\in\zeta$ with a field point 
$z(\tau_1)\in\zeta$; $r_2=-(q\cdot u_2)$. In the forthcoming expressions up 
to the end of the present paper index $1$ indicates that the particle's 
velocity or position is referred to the instant $\tau_1\in ]-\infty,\tau]$ 
while index $2$ says that the particle's characteristics are evaluated at 
instant $\tau_2$ before $\tau$.

Let's explain how these fields appeared. In case we do decide to integrate 
of energy and momentum densities over $\Sigma$ \cite{Yar3D}, we should 
study of interference of outgoing scalar waves emitted by different points 
on particle's world line. This is because massive scalar waves propagate 
not just at a speed of light, but also at all speeds smaller than or equal 
to the speed of light. The world tube is not convenient to study the 
interference. The tilted hyperplane which plays privileged role in the 
radiation reaction problem in classical electrodynamics \cite{Teit} is not 
suitable for our purpose too. The reason is that there is no a plane 
which is orthogonal to the particle's 4-velocities at {\it all} points on 
$\zeta$ before the end point $z(\tau)=\zeta\cap\Sigma$. In 
Refs.\cite{Yar3D,Y3D} we choose the simplest plane $\Sigma_t=\{x\in{\mathbb 
M}_{\,3}: x^0=t\}$ associated with an unmoving inertial observer. 
Non-covariant terms arise unavoidable due to integration over this surface. 
To reveal meaningful contribution in radiated energy-momentum we apply the 
criteria which were first formulated in \cite[Table 1]{Teit}:
\begin{itemize}
\item
the bound term diverges while the radiative one is finite;
\item
the bound component depends on the momentary state of the particle's
motion while the radiative one is accumulated with time; and
\item
the form of the bound terms heavily depends on choosing of an 
integration surface while the radiative terms are invariant. 
\end{itemize}

\begin{figure}[t]
\begin{center}
\epsfclipon
\epsfig{file=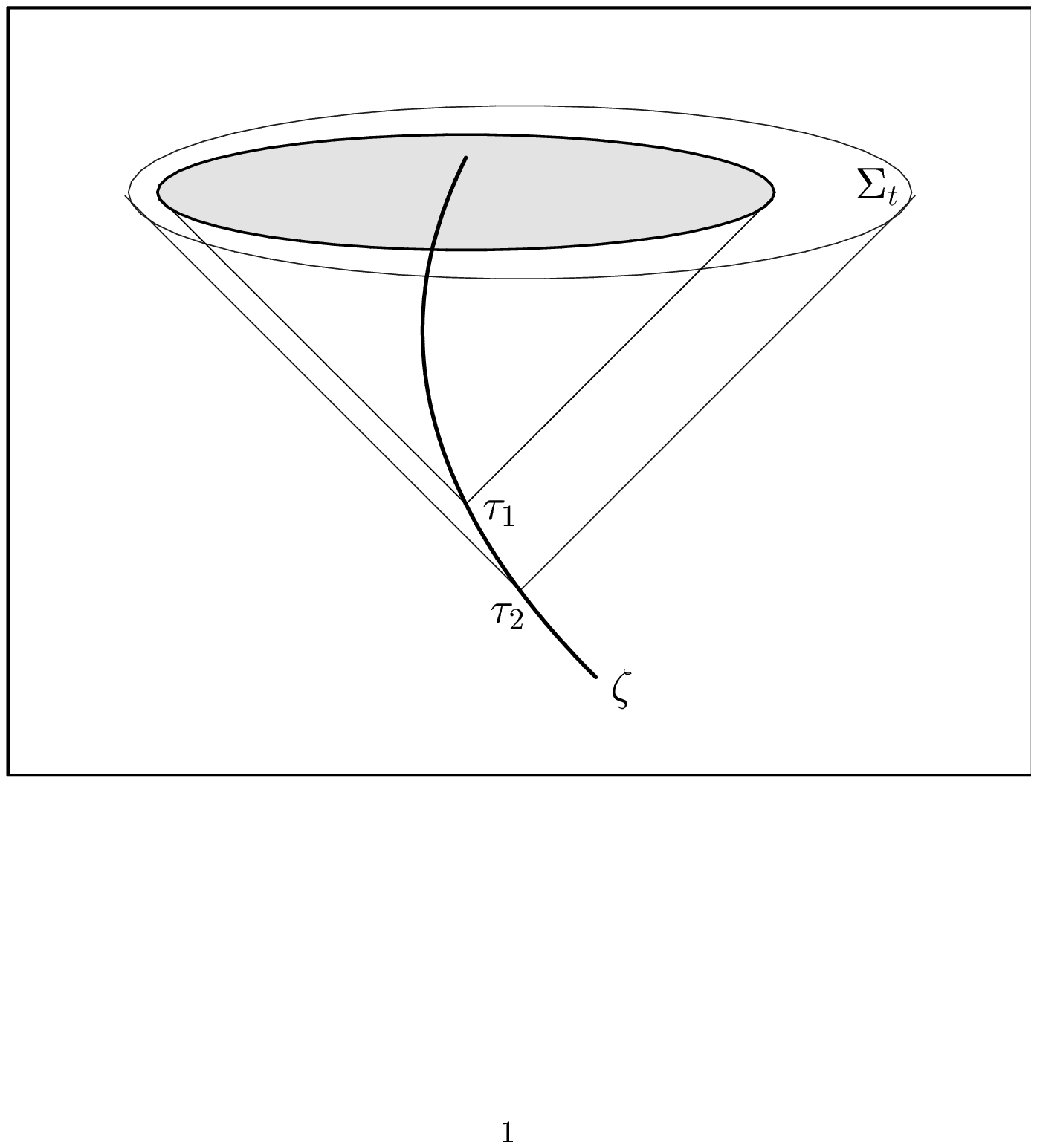,width=0.5\textwidth}$\qquad$
\epsfig{file=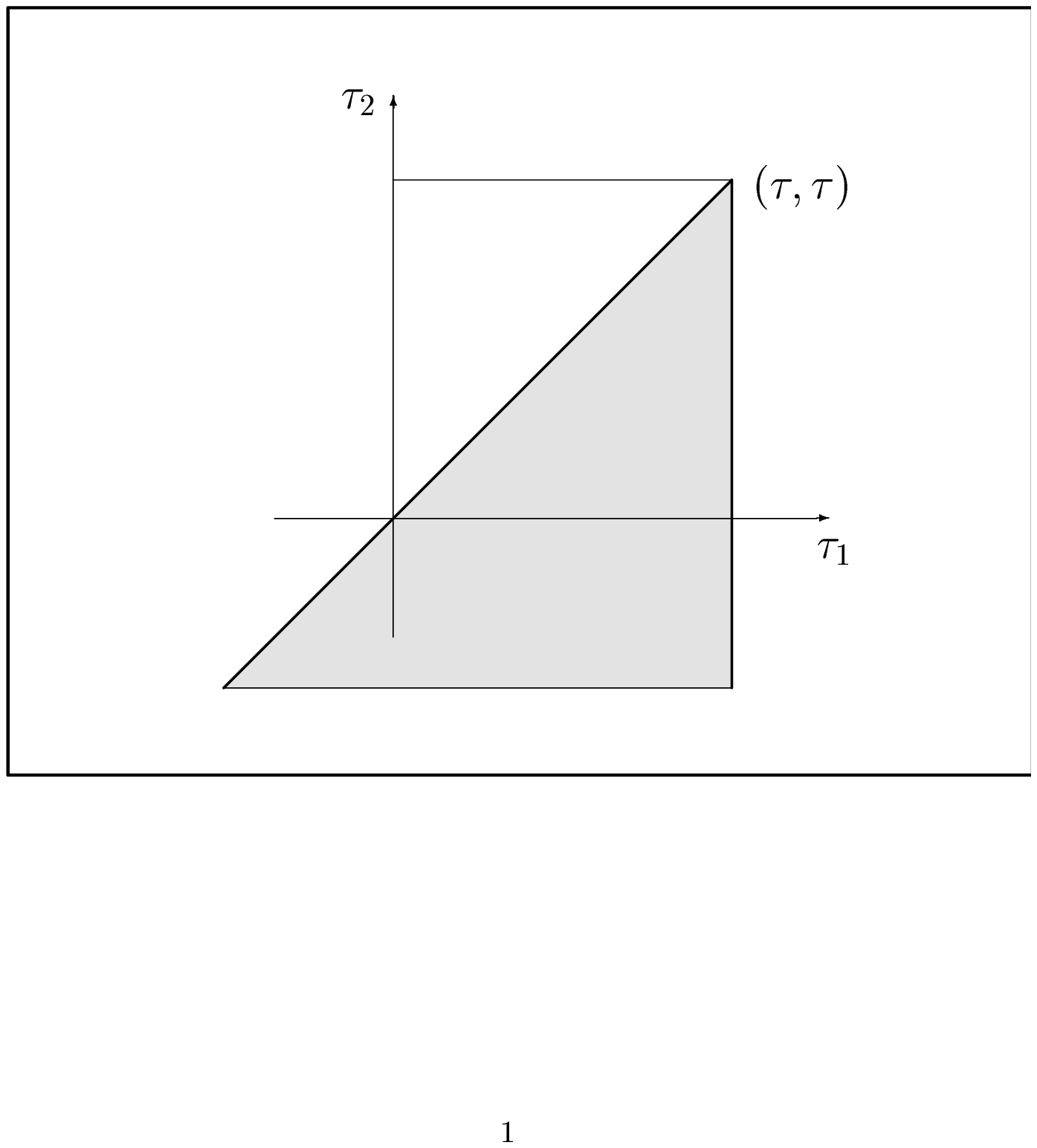,width=0.40\textwidth}
\end{center}
\caption{\label{int_ret}
Outgoing scalar waves generated by the portion of the world line that 
corresponds to the interval $-\infty <\tau_2<\tau_1$ combine within the gray 
disk in the plane $\Sigma_t=\{x\in{\mathbb M}_{\,4}: x^0=t\}$. Corresponding 
domain of integration is $\int_{-\infty}^\tau{\rm 
d}\tau_1\int_{-\infty}^{\tau_1}{\rm d}\tau_2
\int_0^{k_1^0}{\rm d}R\int{\rm d}\Omega$ where $k_1^0$ is the 
radius of smaller disk. After integration over spherical variables the 
domain reduces to double path integral $\int_{-\infty}^\tau{\rm 
d}\tau_1\int_{-\infty}^{\tau_1}{\rm d}\tau_2$. It is sketched on the right 
figure. It is obvious that this domain is equivalent to 
$\int_{-\infty}^\tau{\rm d}\tau_2\int^{\tau}_{\tau_1}{\rm d}\tau_1$.}
\end{figure}

\begin{figure}[t]
\begin{center}
\epsfclipon
\epsfclipon
\epsfig{file=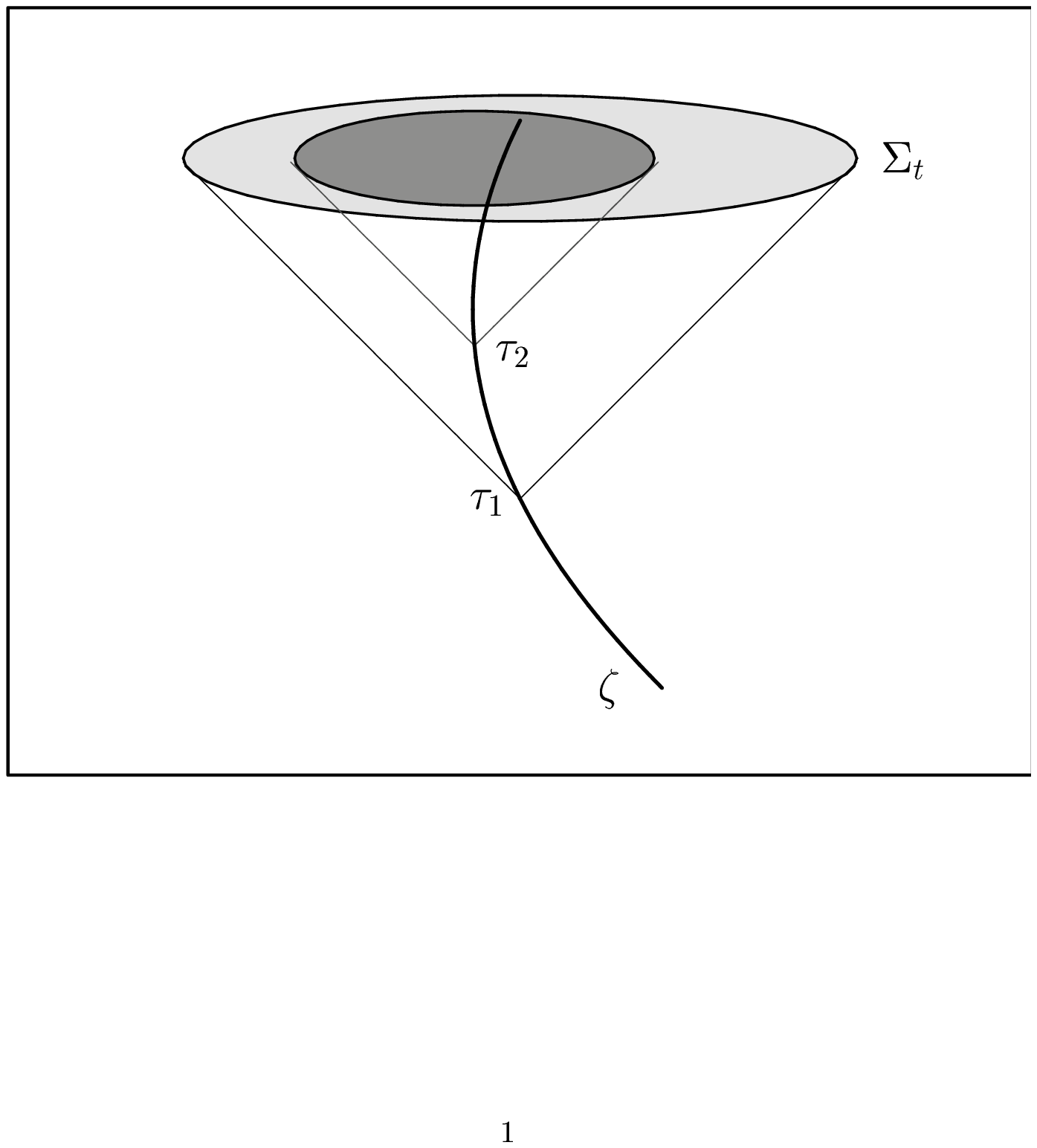,width=0.45\textwidth}$\qquad$
\epsfig{file=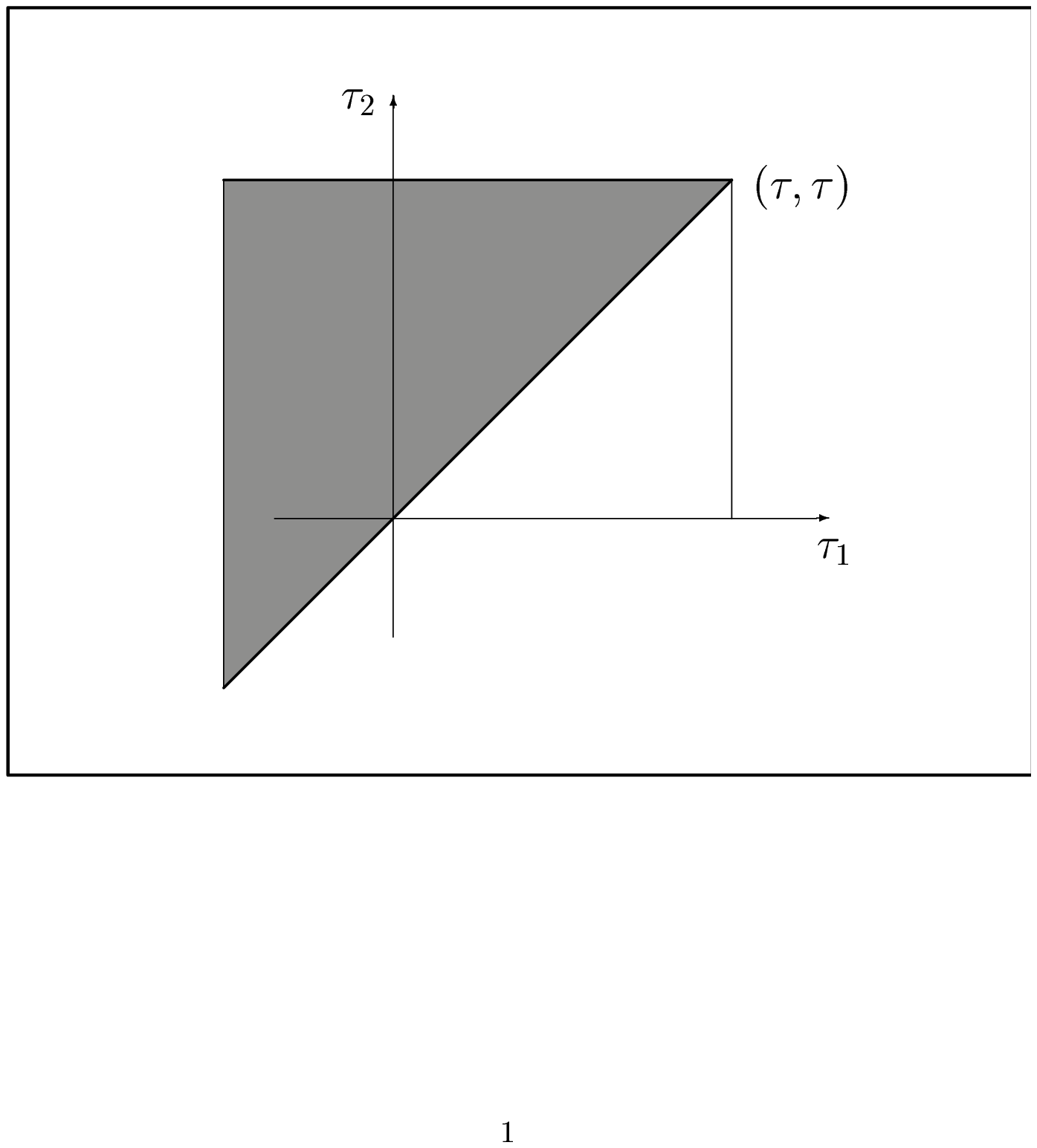,width=0.4\textwidth}
\end{center}
\caption{\label{int_adv}
Outgoing scalar waves generated by the portion of the world line that 
corresponds to the interval $\tau_1<\tau_2\le \tau$  joint together inside 
the dark disk. The domain of integration is $\int_{-\infty}^\tau{\rm 
d}\tau_1\int^{\tau}_{\tau_1}{\rm d}\tau_2
\int_0^{k_2^0}{\rm d}R\int{\rm d}\Omega$ where $k_2^0$ is the 
radius of dark disk. After integration over spherical coordinates which 
parameterize $\Sigma_t$, the double integral $\int_{-\infty}^\tau{\rm 
d}\tau_1\int^{\tau}_{\tau_1}{\rm d}\tau_2$ survives. It can be replaced by 
$\int_{-\infty}^\tau{\rm d}\tau_2\int^{\tau_2}_{-\infty}{\rm d}\tau_1$.}
\end{figure}

Since the stress-energy tensor, either electromagnetic or scalar, is 
quadratic in field strengths, we should {\it twice} integrate it over 
$\zeta$ in order to calculate its flux (\ref{sc-p}) through $\Sigma_t$. 
Figures \ref{int_ret} and \ref{int_adv} picture the interference of the disk 
emanated by fixed point $z(\tau_1)\in\zeta$ with radiation generated by 
portion of the world line from the remote past to the observation instant 
$\tau$. Cumbersome calculations performed in \cite{Yar3D} can be summarized 
as a simple scheme which can be easily adopted to case of massive scalar 
field.

\begin{enumerate}
\item
Integration of the stress-energy tensor density over $\Sigma_t$ yields {\it 
action at a distance} theory which manipulates with fields evaluated on the 
world line only.
\item
In case of combination of waves pictured in Fig.\ref{int_ret} the surface 
integration (\ref{sc-p}) of the stress-energy tensor contributes in {\it 
radiated} energy-momentum one-half of the work 
\begin{equation}\label{pret}
p_{\rm ret}^\mu=-\frac{1}{2}  
\int_{-\infty}^\tau{\rm d}\tau_1 F_{\rm ret}^\mu(\tau_1)
\end{equation}
of the retarded tail force (\ref{tlFret}) acting on the charge itself. (The 
charge ``fills'' its own massive field just as an external one.) 

The surface integration of the angular momentum tensor density (\ref{sc-M}) 
gives one-half of the path integral of the torque of force (\ref{tlFret}):
\begin{equation}\label{Mret}
M_{\rm ret}^{\mu\nu}=-\frac{1}{2}  
\int_{-\infty}^\tau{\rm d}\tau_1 \left[
z_1^\mu F_{\rm ret}^\nu(\tau_1)-z_1^\nu F_{\rm ret}^\mu(\tau_1)
\right].
\end{equation}

\item
Interference of outgoing scalar waves pictured in Fig.\ref{int_adv} 
takes away from {\it radiated} energy-momentum one-half of the work 
\begin{equation}\label{padv}
p_{\rm adv}^\mu=-\frac{1}{2}  
\int_{-\infty}^\tau{\rm d}\tau_1 F_{\rm adv}^\mu(\tau_1)
\end{equation}
of the advanced tail force (\ref{tlFadv}) acting on the charge itself. 
The surface integration of the angular momentum tensor density gives 
one-half of the path integral of the torque of advanced force (\ref{tlFadv}):
\begin{equation}\label{Madv}
M_{\rm adv}^{\mu\nu}=-\frac{1}{2}  
\int_{-\infty}^\tau{\rm d}\tau_1 \left[
z_1^\mu F_{\rm adv}^\nu(\tau_1)-z_1^\nu F_{\rm adv}^\mu(\tau_1)
\right].
\end{equation}

\item
The the support of double integral 
$\int_{-\infty}^\tau {\rm d}\tau_1\int_{\tau_1}^\tau {\rm d}\tau_2$ 
coincides with the support of the integral 
$\int_{-\infty}^\tau {\rm d}\tau_2\int_{-\infty}^{\tau_2}{\rm d}\tau_1$
(see Figs. \ref{int_ret} and \ref{int_adv}). Since instants $\tau_1$ and 
$\tau_2$ label different points at the same world line $\zeta$, one can 
interchanges the indices ``first'' and ``second'' in the integrand. Via 
interchanging of these indices we finally obtain $\int_{-\infty}^\tau {\rm 
d}\tau_1\int_{-\infty}^{\tau_1}{\rm d}\tau_2 g(\tau_2,\tau_1)$
instead of initial 
$\int_{-\infty}^\tau {\rm d}\tau_1\int_{\tau_1}^\tau {\rm d}\tau_2 
g(\tau_1,\tau_2)$.

\item
The radiative part of energy-momentum carried by massive scalar field is 
therefore
\begin{eqnarray}\label{pmR}
p^R_{{\rm tail},\mu}&=-&\frac{1}{2}\int_{-\infty}^\tau {\rm d}\tau_1\left( 
F^{\rm ret}_\mu-F^{\rm adv}_\mu\right)\\
&=&-\frac{g^2}{2}\int_{-\infty}^\tau{\rm 
d}\tau_1\int_{-\infty}^{\tau_1}{\rm 
d}\tau_2\left(f_\mu(z_1,z_2)-f_\mu(z_2,z_1)\right).\nonumber 
\end{eqnarray}
Both the radiated angular momentum
\begin{eqnarray}\label{MmR}
M^R_{{\rm tail},\mu\nu}&=&-\frac{1}{2}\int_{-\infty}^\tau {\rm 
d}\tau_1\left[
z_{1,\mu}\left( F^{\rm ret}_\nu-F^{\rm adv}_\nu\right)
-z_{1,\nu}\left( F^{\rm ret}_\mu-F^{\rm adv}_\mu\right)\right]\\
&=&-\frac{g^2}{2}\int_{-\infty}^\tau{\rm 
d}\tau_1\int_{-\infty}^{\tau_1}{\rm 
d}\tau_2
\left\{\!\!\!\!\phantom{\frac11}
z_{1,\mu}f_\nu(z_1,z_2)-z_{1,\nu}f_\mu(z_1,z_2)\right.
\nonumber\\
&-&\left.\left[z_{2,\mu}f_\nu(z_2,z_1)
-z_{2,\nu}f_\mu(z_2,z_1)\right]\!\!\!\!\phantom{\frac11}\right\}.\nonumber 
\end{eqnarray}
and energy-momentum (\ref{pmR}) exert the radiation reaction.
\end{enumerate}

Now we evaluate the ``upper-limit'' behavior of the tail Noether 
quantities. By this we mean the coincidence limit $\tau_2\to\tau_1$ of the 
of the expressions under the double integrals in eqs. (\ref{pmR}) and 
(\ref{MmR}), namely
\begin{equation}\label{p_nlrad} 
 p_{\rm tail,R}^\mu=-\frac{g^2}{2}\int_{-\infty}^\tau{\rm d}\tau_1
\int_{-\infty}^{\tau_1}{\rm d}\tau_2
k_0^2\frac{J_1(w)}{w}\left[
\frac{1+(q\cdot a_2)}{r_2^2}q^\mu-\frac{u_2^\mu}{r_2}
+\frac{1-(q\cdot a_1)}{r_1^2}q^\mu-\frac{u_1^\mu}{r_1}
\right]
\end{equation} 
and 
\begin{eqnarray}\label{M_nlrad} 
 M_{\rm tail,R}^{\mu\nu}=\frac{g^2}{2}
\int_{-\infty}^\tau{\rm d}\tau_1\int_{-\infty}^{\tau_1}{\rm d}\tau_2
k_0^2\frac{J_1(w)}{w}\left[
\frac{1+(q\cdot 
a_2)}{r_2^2}\left(z_1^\mu z_2^\nu -z_1^\nu z_2^\mu\right)
+\frac{z_1^\mu u_2^\nu-z_1^\nu u_2^\mu}{r_2}
\right.\nonumber\\
\left.
+\frac{1-(q\cdot 
a_1)}{r_1^2}\left(z_1^\mu z_2^\nu -z_1^\nu z_2^\mu\right)
+\frac{z_2^\mu u_1^\nu-z_2^\nu u_1^\mu}{r_1}
\right].
\end{eqnarray}
Let $\tau_1$ be fixed and $\tau_1-\tau_2:=\Delta$ be a small parameter. 
With a degree of accuracy sufficient for our purposes
\begin{eqnarray}\label{delta}
\sqrt{-(q\cdot q)}&=&\Delta\\
q^\mu&=&\Delta\left[u_1^\mu-a_1^\mu\frac{\Delta}{2}+
{\dot a}_1^\mu\frac{\Delta^2}{6}\right]\nonumber\\
u_2^\mu&=&u_1^\mu-a_1^\mu\Delta+{\dot a}_1^\mu\frac{\Delta^2}{2}.\nonumber
\end{eqnarray}
Substituting these into integrands of the double integrals of eqs. 
(\ref{p_nlrad}) and (\ref{M_nlrad}) and passing to the limit $\Delta\to 0$ 
yields vanishing expression. Hence the subscript ``R'' stands for 
``regular'' as well as for ``radiative''.

In the specific case of a uniformly moving source $q^\mu=u^\mu 
(\tau_1-\tau_2)$ and $r_1=r_2=\tau_1-\tau_2$. It immediately gives 
$f_\mu(z_1,z_2)=f_\mu(z_2,z_1)$ and, therefore, the integrands in eqs. 
(\ref{p_nlrad}) and (\ref{M_nlrad}) are identically equal to zero. The 
local parts of radiation (\ref{p_sc}) and (\ref{M_sc}) vanish if 
$u^\mu={\rm const}$. As could be expected, nonaccelerating scalar charge 
does not radiate.

In the following Section we check the formulae (\ref{pmR}) and (\ref{MmR})
via analysis of energy-momentum and angular momentum balance equations. 
Analogous equations yield correct equation of motion of radiating charge in 
conventional electrodynamics \cite{Y4D} as well as in flat spacetime of six 
dimensions \cite{Y6D}. It is reasonable to expect that conservation laws 
result correct equation of motion of point-like source coupled with massive 
scalar field where radiation back reaction is taken into account. 

\section{Balance equations}\label{meq}
\setcounter{equation}{0}

The equation of motion of radiating pole of massive scalar field was derived 
by Harish-Chandra \cite{HC} in 1946. (An alternative derivation was produced 
by Havas and Crownfield in \cite{HvC}.) Following the method of Dirac 
\cite{Dir}, Harish-Chandra enclosed the world line of the particle by a 
narrow tube, the radius of which will in the end be made to tend to zero. 
The author calculates the flow of energy and momentum out of the portion of 
the tube in presence of an external field. The condition was imposed that 
the flow depends only on the states at the two ends of the tube (the 
so-called ``inflow theorem'', see \cite{BHC44,BHC}). After integration over 
the tube along the world line and a limiting procedure, the equation of 
motion was derived. In our notation it looks as follows:
\begin{eqnarray}\label{HCme}
&&\!\!\!\!\!\!\!\!\!\!m_0a_\tau^\mu-\frac{g^2}{3}\left({\dot a}^\mu_\tau
-a^2_\tau u^\mu_\tau\right)-\frac{g^2}{2}k_0^2u_\tau^\mu
+g^2\int_{-\infty}^\tau {\rm d} sk_0^4\frac{J_2(w)}{w^2}q^\mu
+g^2\frac{{\rm d}}{{\rm d}\tau}\left(u_\tau^\mu\int_{-\infty}^\tau 
{\rm d} sk_0^2\frac{J_1(w)}{w}\right)\nonumber\\
&=&g\eta^{\mu\alpha}\frac{\partial \varphi_{\rm ext}}{\partial 
z^\alpha}+g\frac{{\rm d}}{{\rm d}\tau}\left(u_\tau^\mu \varphi_{\rm 
ext}\right)
\end{eqnarray}
where $m_0$ is an arbitrary constant identified with the mass of the 
particle and $\varphi_{\rm ext}$ is the scalar potential of the external 
field evaluated at the current position of the particle. $J_2(w)$ is the 
second order Bessel's function of $w=k_0\sqrt{-(q\cdot q)}$. In this 
Section the Harish-Chandra equation will be obtained via analysis of 
energy-momentum and angular momentum balance equations.

In previous Section we introduce the radiative part $p_{\rm R}=p_{\rm 
dir,R}+p_{\rm tail,R}$ of energy-momentum carried by the field. It defines 
loss of energy and momentum due to scalar radiation. The bound part, $p_{\rm 
S}$, is absorbed by particle's 4-momentum so that dressed scalar charge 
would not undergo any additional radiation reaction. Already renormalized 
particle's individual four-momentum, say $p_{\rm part}$, together with 
$p_{\rm R}$ constitute the total energy-momentum of our composite particle 
plus field system:  $P=p_{\rm part}+p_{\rm R}$. We suppose that the external 
scalar force $g\Phi_{\rm ext}$ matches the change of $P$ with time: 
\begin{equation}\label{blp-sc}
{\dot p}^\mu_{\rm part}+{\dot p}^\mu_{\rm R}=g\Phi^\mu_{\rm ext}.
\end{equation}
The overdot indicates differentiation with respect to proper time 
parameter $\tau$.

Differentiating the tail contribution (\ref{p_nlrad}) in radiated 
energy-momentum, we obtain the single path integral:
\begin{equation}\label{dotp-sc}
{\dot p}^\mu_{\rm tail}(\tau)=-\frac{g^2}{2}\int_{-\infty}^\tau{\rm d} s
k_0^2\frac{J_1(w)}{w}\left[
\frac{1+(q\cdot a_s)}{r_s^2}q^\mu-\frac{u_s^\mu}{r_s}+\frac{1-(q\cdot 
a_\tau)}{r_\tau^2}q^\mu-\frac{u_\tau^\mu}{r_\tau}\right].
\end{equation}
Here index $\tau$ indicates that particle's position, velocity, or 
acceleration is referred to the observation instant $\tau$ while index $s$ 
says that the particle's characteristics are evaluated at instant 
$s\le\tau$. The expression should be added to the proper time derivative 
of energy-momentum (\ref{p_sc}) carried by a massless scalar field. 
Substituting the sum ${\dot p}_{\rm dir}+{\dot p}_{\rm tail}$ for ${\dot 
p}_{\rm R}$ into eq.(\ref{blp-sc}), we obtain the energy-momentum balance 
equation:
\begin{eqnarray}\label{pdot-sc}
{\dot p}^\mu_{\rm part}(\tau)&=&-\frac{g^2}{3}a^2(\tau)u^\mu_\tau 
+\frac{g^2}{2}\int_{-\infty}^\tau{\rm d} s
k_0^2\frac{J_1(w)}{w}\left[
\frac{1+(q\cdot a_s)}{r_s^2}q^\mu-\frac{u_s^\mu}{r_s}
+\frac{1-(q\cdot 
a_\tau)}{r_\tau^2}q^\mu\right.\nonumber\\
&-&\left.\frac{u_\tau^\mu}{r_\tau}\right]
+g\Phi^\mu_{\rm ext}.
\end{eqnarray} 
Our next task is to derive expression which explain how four-momentum of 
``dressed'' scalar charge depends on its individual characteristics 
(velocity, position, mass etc.). 

We do not make any assumptions about the particle structure, its charge 
distribution and its size. We only assume that the particle four-momentum 
$p_{\rm part}$ is finite. To find out the desired expression we analyze 
conserved quantities corresponding to the invariance of the theory under 
proper homogeneous Lorentz transformations. The total angular momentum, say 
$M$, consists of particle's angular momentum $z\wedge p_{\rm part}$ and 
radiative part of angular momentum carried by massive scalar field:
\begin{equation}\label{Mtot-sc}
M^{\mu\nu}=z_\tau^\mu p_{\rm part}^\nu(\tau) 
- z_\tau^\nu p_{\rm part}^\mu(\tau) + M^{\mu\nu}_{\rm R}(\tau).
\end{equation}
The radiated angular momentum $M^{\mu\nu}_{\rm R}=M^{\mu\nu}_{\rm R,dir}+ 
M^{\mu\nu}_{\rm R,tail}$ is determined by eqs. (\ref{M_sc}) 
(direct part) and (\ref{M_nlrad}) (tail part), respectively. We assume that 
the torque $g(z_\tau^\mu\Phi^\nu_{\rm ext}-z_\tau^\nu\Phi^\mu_{\rm ext})$ of 
the external force matches the change of $M$ with time. Having 
differentiated the angular momentum expression (\ref{Mtot-sc}) and inserting 
eq.(\ref{pdot-sc}), we arrive at the equality
\begin{equation}\label{Mdot-sc}
u_\tau\wedge \left(p_{\rm part}+\frac{g^2}{3}a_\tau+
\frac{g^2}{2}\int_{-\infty}^\tau {\rm d} sk_0^2\frac{J_1(w)}{w}
\frac{q}{r_\tau}\right)=0,
\end{equation}
where $r_\tau=-(q\cdot u(\tau))$. Symbol $\wedge$ denotes the wedge product.

Apart from usual velocity term, the 4-momentum of dressed scalar charge
contains also a contribution from field:
\begin{equation}\label{ppart}
p_{\rm part}^\mu=mu_\tau^\mu-\frac{g^2}{3}a_\tau^\mu-
\frac{g^2}{2}\int_{-\infty}^\tau {\rm d} sk_0^2\frac{J_1(w)}{w}
\frac{q^\mu}{r_\tau}.
\end{equation}
The local part is the scalar analog of Teitelboim's expression \cite{Teit} 
for individual 4-momen\-tum of a dressed electric charge in classical. The 
tail term is then nothing but the bound part of energy-momentum carried by 
the massive scalar field:
\begin{equation}\label{pmS}
p_{\rm tail,S}^\mu=-\frac12\int_{-\infty}^\tau{\rm d}\tau_1\left(F_{\rm 
ret}^\mu+F_{\rm adv}^\mu\right).
\end{equation}
The bound part of the field energy-momentum is permanently ``attached'' to 
the charge and is carried along with it. It is worth noting that the 
expression under the integral sign does not diverge if $s\to\tau$. The 
``local'' Coulomb-like infinity is the only divergency stemming from the 
pointness of the source (see \ref{unmvd}).

In the specific case uniform motion $u^\mu={\rm const}$ and argument of 
Bessel's function simplifies: $w=k_0(\tau-s)$. Similarly 
$q^\mu/r_\tau=u^\mu$. Since
\begin{equation}\label{b_int}
\int_{-\infty}^\tau{\rm d} s\frac{J_1[k_0(\tau -s)]}{\tau -s}=1,
\end{equation}
the field generated by a uniformly moving charge contributes an amount 
$p^\mu_{\rm tail,S}=-1/2g^2k_0u^\mu$ to its energy-momentum. This finding is 
in line with that of \ref{unmvd} where is established that if the 
particle is permanently at rest, the scalar meson field adds  $-1/2g^2k_0$ 
to its energy.

The expression for the scalar function $m(\tau)$ is find in \ref{mass} via 
analysis of differential consequences of conservation laws. We derive that 
{\it already renormalized} dynamical mass $m$ depends on particle's 
evolution before the observation instant $\tau$:
\begin{equation}\label{mm}
m=m_0+g^2\int_{-\infty}^\tau {\rm d} 
sk_0^2\frac{J_1[w(\tau,s)]}{w(\tau,s)}-
g\varphi_{\rm ext}.
\end{equation} 
The constant $m_0$ can be identified with the renormalization constant 
in action (\ref{Itot}) which governs the dynamics of a point-like charge 
coupled to massive scalar field. $m_0$ absorbs Coulomb-like divergence 
stemming from local part of potential (\ref{sg-sm}). It is of great 
importance that the dynamical mass, $m$, will vary with time: the particle 
will necessarily gain or lost its mass as a result of interactions with its 
own field as well as with the external one. The field of a uniformly moving 
charge contributes an amount $g^2k_0$ to its inertial mass.

To derive the effective equation of motion of radiating charge we replace 
${\dot p}_{\rm part}^\mu$ in left-hand side of eq.(\ref{pdot-sc}) by 
differential consequence of eq.(\ref{ppart}). We apply the formula
\begin{equation}\label{df}
\frac{\partial}{\partial\tau}\int_{-\infty}^\tau{\rm d} sf(\tau,s)=
\int_{-\infty}^\tau{\rm d} s\left(\frac{\partial f}{\partial\tau}+
\frac{\partial f}{\partial s}\right).
\end{equation} 
At the end of a straightforward calculations, we obtain 
\begin{equation}\label{me}
 ma^\mu_\tau+{\dot m}u^\mu_\tau=\frac{g^2}{3}\left({\dot a}^\mu_\tau
-a^2_\tau u^\mu_\tau\right)
+ g^2\int_{-\infty}^\tau {\rm d} sk_0^2\frac{J_1(w)}{w}\left[
\frac{1+(q\cdot a_s)}{r_s^2}q^\mu-\frac{u_s^\mu}{r_s}
\right]+g\Phi^\mu_{\rm ext},
\end{equation}
where dynamical mass $m(\tau)$ is defined by eq.(\ref{mm}). The direct
part of the self-force is one-half of well-known Abraham radiation reaction 
vector, while the tail one is then nothing but the tail part of 
particle's scalar field strengths (\ref{Fmret}) acting upon itself.

Now we compare this effective equation of motion with the Harish-Chandra 
equation (\ref{HCme}). The latter can be simplified substantially.
Having used the recurrent relation
\begin{equation}\label{rcJ}
J_2(w)=\frac{J_1(w)}{w}-\frac{{\rm d} J_1(w)}{{\rm d}w}
\end{equation}
between Bessel functions of order two and of order one, after integration by 
parts we obtain
\begin{equation}\label{rc}
g^2\int_{-\infty}^\tau {\rm d} sk_0^4\frac{J_2(w)}{w^2}q^\mu=
\frac{g^2}{2}k_0^2u_\tau^\mu -
g^2\int_{-\infty}^\tau {\rm d} sk_0^2\frac{J_1(w)}{w}\left[
\frac{1+(q\cdot a_s)}{r_s^2}q^\mu-\frac{u_s^\mu}{r_s}
\right].
\end{equation}
We also collect all the total time derivatives involved in Harish-Chandra 
equation (\ref{HCme}). The term $m(\tau)u_\tau^\mu$ arises under the time 
derivative operator, where time-dependent function $m(\tau)$ is then nothing 
but the dynamical mass (\ref{mm}) of the particle. On rearrangement, the 
Harish-Chandra equation of motion (\ref{HCme}) coincides with the equation 
(\ref{me}) which is obtained via analysis of balance equations. It is in 
favor of the renormalization scheme for tail theories developed in the
present paper.

To clear physical sense of the effective equation of motion (\ref{me}) we 
move the velocity term ${\dot m}u_\tau^\mu$ to the right-hand side of this 
equation:
\begin{equation}\label{meK}
m(\tau)a_\tau^\mu=\frac{g^2}{3}\left({\dot a}^\mu_\tau
-a^2_\tau u^\mu_\tau\right) + f_{\rm self}^\mu + f_{\rm ext}^\mu.
\end{equation}
According to \cite{Kos}, the scalar potential produces the Minkowski force
\begin{equation}
f_{\rm ext}^\mu = g\left(\eta^{\mu\alpha} + u_\tau^\mu 
u_\tau^\alpha\right)\frac{\partial\varphi_{\rm ext}}{\partial z^\alpha}
\end{equation}
which is orthogonal to the particle's 4-velocity. The self-force
\begin{equation}
 f_{\rm self}^\mu = g^2\int_{-\infty}^\tau {\rm d} sk_0^2
\frac{J_1(w)}{w}\left[
\frac{1+(q\cdot a_s)}{r_s^2}\left(q^\mu-r_\tau u_\tau^\mu\right) 
-\frac{u_s^\mu+(u_s\cdot u_\tau)u_\tau^\mu}{r_s}
\right]
\end{equation}
is constructed analogously from the tail part of gradient (\ref{Fmret}) of 
particle's potential (\ref{sg-sm}) supported on the world line $\zeta$. 
Indeed, since the massive scalar field propagates at all speeds smaller 
than the speed of light, the charge may ``fill'' its own field, which will 
act on it just like an external field. The own field contributes also to 
particle's inertial mass $m(\tau)$ defined by eq.(\ref{mm}).

\section{Conclusions}\label{concl}
\begin{figure}[t]
\begin{center}
\epsfclipon
\epsfig{file=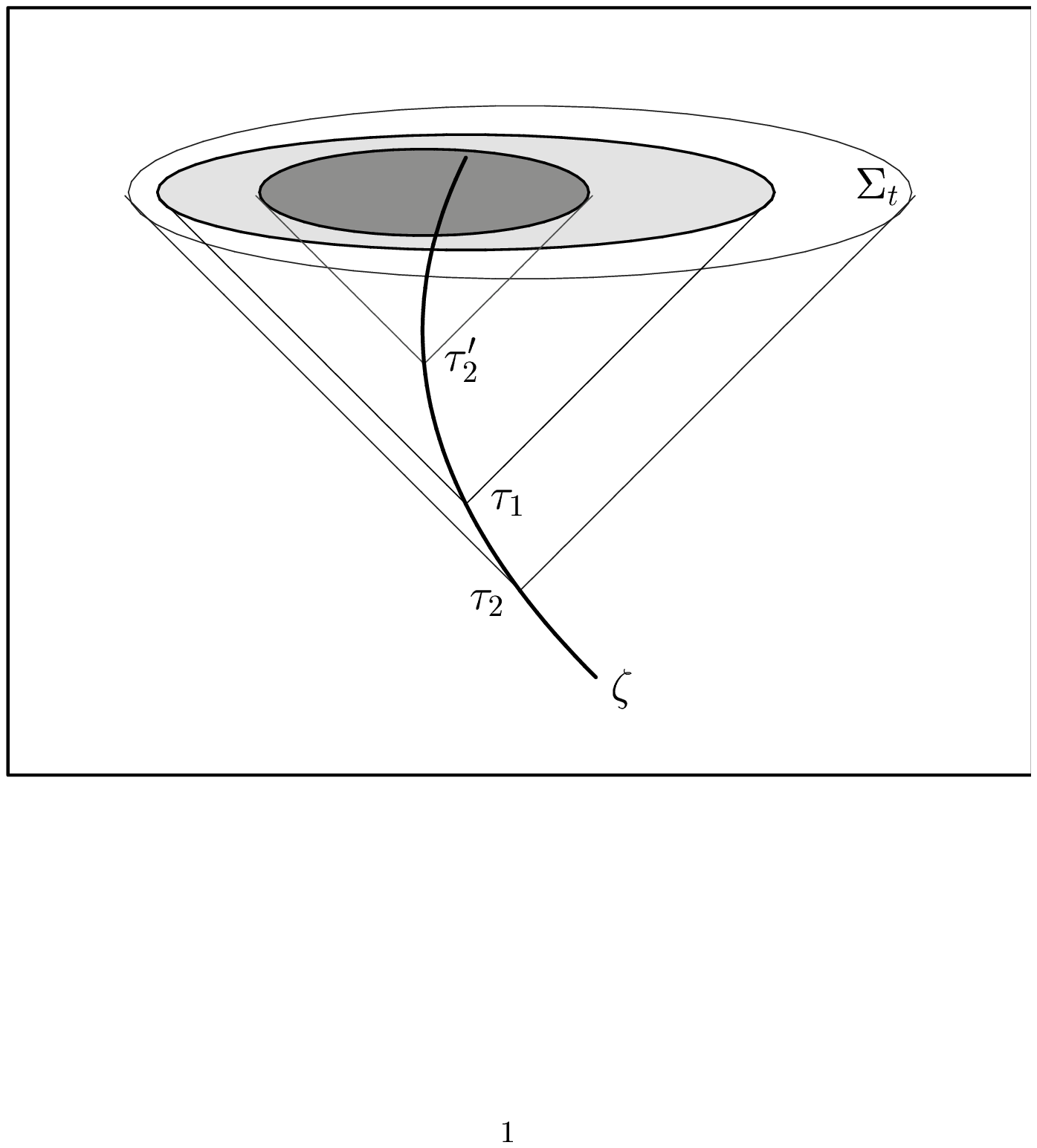,width=0.43\textwidth}
\end{center}
\caption{\label{int_ra}
We call ``retarded'' the force (\ref{tlFret}) with integration over the 
portion of the world line {\it before} $\tau_1$. We call ``advanced'' the 
force (\ref{tlFadv}) with integration over the portion of the world line 
{\it after} $\tau_1$. For an observer placed at point $z(\tau_1)\in\zeta$ 
the radiative (\ref{pmR}) and the bound (\ref{pmS}) parts of massive 
scalar field momentum looks as the combination of incoming and outgoing 
radiation, and yet the retarded causality is not violated. We still 
consider the interference of outgoing waves presented at the observation 
instant $\tau$. The scalar field carries information about the charge's 
past.}
\end{figure}

In the present paper, we adopt the Dirac scheme of decomposition of the 
retarded Green's function into symmetric (singular) and radiative (regular) 
parts to functions supported within light cones. The regularization scheme 
summarizes a scrupulous analysis of energy-momentum and angular momentum 
balance equations in $2+1$ electrodynamics \cite{Yar3D,Y3D}. It differs 
from the approach developed by Detweiler and Whiting \cite{DW} on two 
``extra'' entities: additional instant $\tau_1$ before the instant of 
observation and extra integration of the retarded and the advanced tail 
forces over particle's path. So, the retarded tail force depends on 
the particle's past history before $\tau_1<\tau$ (see Figure \ref{int_ra}). 
Its advanced counterpart is generated by portion of the world line that 
corresponds to the interval $[\tau_1,\tau]$. The tail part of radiated 
energy-momentum is one-half of the work done by the retarded force minus 
one-half of the work done by the advanced force, taken with opposite sign. 
This part of radiation detaches the point source and leads an independent 
existence.

The support of the advanced force is a portion of the world line that 
corresponds to finite time interval. It is presented implicitly in the 
Detweiler and Whiting construction where the tail term (\ref{DWadd}) is 
introduced to achieve appropriate retarded causality. It ``evaporates'' in 
the coincidence limit $x\to z(\tau)$ at which its support $[\tau^{\rm 
ret}(x),\tau^{\rm adv}(x)]$ shrinks to zero.

The one-half sum of the retarded and the advanced works is the bound 
part of tail energy-momentum which is permanently attached to the charge 
and is carried along with it. It modifies particle's individual
characteristics (its momentum and its inertial mass). A point source 
together with surrounded ``cloud'' constitute dressed charged particle.

Since the properties of the retarded and the advanced solutions of wave 
equation, the one-half sum is singular while one-half difference is regular
at the location of the particle.

The bound and the radiative angular momentum carried by charge's  
field are simply torques of the above combinations of the retarded and the 
advanced tail forces.

Together with contributions from the direct part of Green's function, the 
tail terms constitute Noether quantities removing by outgoing waves from 
the dressed source. Chan\-ges in individual momentum and angular momentum 
of dressed charged particle compensate losses of energy, momentum, and 
angular momentum due to radiation. (Influence of an external device can be 
modelled easily.) Analysis of balance equations yields the Harish-Chandra 
equation of motion of radiating scalar pole \cite{HC}. This equation 
includes the effect of particle's own field as well as the influence of an 
external force.

Energy-momentum and angular momentum balance equations for radiating scalar 
pole constitute system of ten linear algebraic equations in variables 
$p^\mu_{\rm part}(\tau)$ and their first time derivatives ${\dot p}^\mu_{\rm 
part}(\tau)$ as functions of particle's individual characteristics 
(velocity, acceleration, charge etc.). The system is degenerate, so that 
solution for particle's 4-momentum includes arbitrary scalar function, 
$m(\tau)$, which can be identified with the dynamical mass of the particle. 
Besides renormalization constant, the mass includes contributions from 
particle's own field as well as from an external field. 

This is a special feature of the self force problem for a scalar charge. 
Indeed, the time-varying mass arises also in the radiation reaction for a 
pointlike particle coupled to a massless scalar field on a curved background 
\cite{Q}. The phenomenon of mass loss by scalar charge is studied in 
\cite{BHPs,HPs}. Similar phenomenon occurs in the theory which describe a 
point-like charge coupled with massless scalar field in flat spacetime of 
three dimensions \cite{B}. The charge loses its mass through the emission of 
monopole radiation.  

\section*{Acknowledgments}
I am grateful to V.Tre\-tyak for continuous encouragement and for 
a helpful reading of this manuscript. I would like to thank A.Du\-vi\-ryak 
for many useful discussions.

\subsubsection{Energy-momentum of the scalar massive field of uniformly 
moving source}\label{unmvd}
\setcounter{equation}{0}

In this Appendix we calculate the energy-momentum (\ref{sc-p})
carried by the scalar massive field due to static charge $g$. The 
stress-energy tensor $\hat T$ is given by eq.(\ref{T}) where $\varphi$ is 
Yukawa potential (\ref{YuP}).

It is convenient to choose the simplest plane $\Sigma_t=\{x\in{\mathbb 
M}_{\,4}:x^0=t\}$ associated with unmoving observer. We start with the 
spherical coordinates
\begin{equation}
x^0=s+r,\qquad x^i=rn^i
\end{equation}
where $n^i=(\cos\phi\sin\theta,\sin\phi\sin\theta,\cos\theta)$ and $s$ is 
the parameter of evolution. To adopt them to the integration surface 
$\Sigma_t$ we replace the radius $r$ by the expression $t-s$. On 
rearrangement, the final coordinate transformation 
$(x^0,x^1,x^2,x^3)\mapsto (t,s,\phi,\theta)$ looks as follows:
\begin{equation}
x^0=t,\qquad x^i=(t-s)n^i.
\end{equation}
The surface element is given by
\begin{equation}
{\rm d}\sigma_0=(t-s)^2{\rm d} s{\rm d}\Omega
\end{equation}
where ${\rm d}\Omega=\sin\theta{\rm d}\theta{\rm d}\phi$ is an element of 
solid angle.

After trivial calculation one can derive the only non-trivial component of 
energy-momentum (\ref{sc-p}) is 
\begin{eqnarray}
p^0_{\rm sc}&=&\frac{1}{4\pi}
\int_{-\infty}^t{\rm d} s(t-s)^2\int{\rm d}\Omega\frac12\left[
\sum_i\left(\frac{\partial\varphi}{\partial x^i}\right)^2+k_0^2\varphi^2
\right]
\\
&=&\frac{g^2}{2}\left[
k_0\exp[-2k_0(t-s)]+\frac{\exp[-2k_0(t-s)]}{t-s}
\right]_{s\to-\infty}^{s\to t}
\nonumber\\
&=&\lim_{\varepsilon\to 0}\frac{g^2}{2\varepsilon} - \frac{g^2}{2}k_0
\nonumber
\end{eqnarray}
where $\varepsilon$ is positively valued small parameter. 

Having performed Poincar\'e transformation, the combination of 
translation and Lorentz transformation, we find the energy-momentum 
carried by massive scalar field of uniformly moving charge:
\begin{equation}\label{infin}
p^\mu_{\rm sc}=\lim_{\varepsilon\to 0}\frac{g^2}{2\varepsilon}u^\mu - 
\frac{g^2}{2}k_0u^\mu .
\end{equation}
The divergent Coulomb-like term is absorbed by the ``bare'' mass $m_0$ 
involved in action integral (\ref{Itot}) while the finite term contributes 
to the particle's individual 4-momentum (\ref{ppart}).

\subsubsection{Derivation of the renormalized mass of scalar 
charge}\label{mass}
\setcounter{equation}{0}
We follow the scheme elaborated within analysis of $2+1$ electrodynamics 
where the expression for the scalar function $m(\tau)$ is derived via 
analysis of differential consequences of conservation laws. In 
hypothetical Minkowski space of three dimensions {\it already renormalized} 
mass of charged particle depends on particle's evolution before the 
observation instant $\tau$.

The scalar product of particle 4-velocity on the time derivative of 
particle 4-momentum (\ref{pdot-sc}) is as follows:
\begin{eqnarray}\label{udp}
({\dot p}_{\rm part}\cdot u_\tau)&=&\frac{g^2}{3}a_\tau^2+
\frac{g^2}{2}\int_{-\infty}^\tau {\rm d} s
k_0^2\frac{J_1(w)}{w}\left[
\frac{1+(q\cdot a_s)}{r_s^2}(q\cdot u_\tau)-\frac{(u_s\cdot u_\tau)}{r_s}+
\frac{(q\cdot a_\tau)}{r_\tau}
\right]\nonumber\\
&+&g(u\cdot\Phi_{\rm ext}).
\end{eqnarray}
Since $(u\cdot a)=0$, the scalar product of particle acceleration on the 
particle 4-momentum (\ref{ppart}) does not contain the scalar function 
$m$:
\begin{equation}\label{ap}
(p_{\rm part}\cdot a_\tau)=-\frac{g^2}{3}a_\tau^2-
\frac{g^2}{2}\int_{-\infty}^\tau {\rm d} s
k_0^2\frac{J_1(w)}{w}\frac{(q\cdot a_\tau)}{r_\tau}.
\end{equation}
Summing up (\ref{udp}) and (\ref{ap}) we obtain
\begin{equation}\label{d2p}
\frac{{\rm d}}{{\rm d}\tau}(p_{\rm part}\cdot u_\tau)=
\frac{g^2}{2}\int_{-\infty}^\tau {\rm d} s
k_0^2\frac{J_1(w)}{w}
\frac{\partial}{\partial s}\left[\frac{(q\cdot u_\tau)}{r_s}\right]
+g(u\cdot\Phi_{\rm ext}).
\end{equation}
We rewrite the expression under the integral sign as the following 
combination of partial derivatives in time variables:
\begin{equation}
k_0^2\frac{J_1(w)}{w}
\frac{\partial}{\partial s}\left(\frac{(q\cdot u_\tau)}{r_s}\right)=
\frac{\partial}{\partial s}\left(k_0^2\frac{J_1(w)}{w}
\frac{(q\cdot u_\tau)}{r_s}\right)-
\frac{\partial}{\partial \tau}\left(k_0^2\frac{J_1(w)}{w}\right).
\end{equation}
At this point we suppose that the external field is the gradient of an 
external scalar potential, say $\varphi_{\rm ext}(x)$. If the field is 
referred to the point $z(\tau)$ where the charge is located, the scalar 
product $(u\cdot\Phi_{\rm ext})$ becomes the total time derivative:
\begin{eqnarray}
(u\cdot\Phi_{\rm ext})&=&\frac{{\rm d}z^\alpha}{{\rm d}\tau}
\frac{\partial \varphi_{\rm ext}(z)}{\partial z^\alpha}\\
&=&\frac{{\rm d}\varphi_{\rm ext}}{{\rm d}\tau}.\nonumber
\end{eqnarray}
These circumstances allow us to integrate the expression (\ref{d2p}) over
$\tau$:
\begin{eqnarray}
(p_{\rm part}\cdot 
u_\tau)&=&-m_0+\frac{g^2}{2}\left.
\begin{array}{c}
\displaystyle 
\int_{-\infty}^\tau{\rm d}\tau_1\int_{-\infty}^{\tau_1}{\rm d}\tau_2\\
\\[-1em]
\displaystyle
\int_{-\infty}^\tau{\rm d}\tau_2\int_{\tau_2}^\tau{\rm d}\tau_1
\end{array}
\right\}\left[
\frac{\partial}{\partial \tau_2}\left(k_0^2\frac{J_1(w)}{w}
\frac{(q\cdot u_1)}{r_2}\right)\right.\nonumber\\
&-&\left.
\frac{\partial}{\partial \tau_1}\left(k_0^2\frac{J_1(w)}{w}\right)
\right]+g\varphi_{\rm ext}\nonumber\\
&=&-m_0-\frac{g^2}{2}\int_{-\infty}^\tau{\rm d} s
k_0^2\frac{J_1(w(\tau,s))}{w(\tau,s)}
+g\varphi_{\rm ext}.
\end{eqnarray}
The external potential is referred to the particle's position $z(\tau)$.

Alternatively, the scalar product of 4-momentum (\ref{ppart}) and 
4-velocity is as follows:
\begin{equation}\label{pu}
(p_{\rm part}\cdot u_\tau)=-m+\frac{g^2}{2}\int_{-\infty}^\tau ds
k_0^2\frac{J_1[w(\tau,s)]}{w(\tau,s)}.
\end{equation} 
Having compared these expressions we obtain:
\begin{equation}\label{mas-sq}
m=m_0+g^2\int_{-\infty}^\tau {\rm d} sk_0^2\frac{J_1[w(\tau,s)]}{w(\tau,s)}-
g\varphi_{\rm ext}[z(\tau)].
\end{equation} 
The constant $m_0$ can be identified with the renormalization constant 
in action (\ref{Itot}) which governs the dynamics of a point-like charge 
coupled to massive scalar field. $m_0$ absorbs Coulomb-like divergence 
(\ref{infin}) stemming from direct part of potential (\ref{sg-sm}). It is 
of great importance that the dynamical mass, $m$, will vary with time: the 
particle will necessarily gain or lost its mass as a result of interactions 
with its own field as well as with the external one. Since 
eq.(\ref{b_int}), the field of a uniformly moving charge contributes an 
amount $g^2k_0$ to its inertial mass.


\begin{thebibliography}{99}
\bibitem{WB}
B. S. DeWitt and R. W. Brehme, Ann.Phys. (N.Y.) {\bf 9}, 220 (1960).

\bibitem{Hb}
J. M. Hobbs, Ann.Phys. (N.Y.) {\bf 47}, 141 (1968).

\bibitem{MST}
Y. Mino, M. Sasaki, and T. Tanaka, Phys. Rev. D {\bf 55}, 3457 (1997).

\bibitem{Q}
T. C. Quinn, Phys. Rev. D {\bf 62}, 064029 (2000).

\bibitem{Pois}
E. Poisson, Living Rev. Relativity {\bf 7}, Irr-2004-6 (2004).
arXiv:gr-qc/0306052 

\bibitem{DW}
S. Detweiler and B.F. Whiting, Phys. Rev. D {\bf 67}, 024025 (2003).

\bibitem{Dir}
P.A.M. Dirac, Proc. R. Soc. A (London) {\bf 167}, 148 (1938).

\bibitem{Teit}
C. Teitelboim, Phys. Rev. D {\bf 1}, 1572 (1970).

\bibitem{Yar3D}
Yu. Yaremko, J. Math. Phys. {\bf 48}, 092901 (2007).

\bibitem{Y3D}
Yu. Yaremko, J.Phys.A: Math.Theor. {\bf 40}, 13161 (2007).

\bibitem{Gl}
D. V. Gal'tsov, Phys. Rev. D {\bf 66}, 025016 (2002).

\bibitem{KLS}
P. O. Kazinski, S. L. Lyakhovich, and A. A. Sharapov, Phys. Rev. D {\bf 
66}, 025017 (2002).

\bibitem{HC}
Harish-Chandra, Proc. R. Soc. (London) {\bf A185}, 269 (1946).

\bibitem{Kos}
B. Kosyakov, {\it Introduction to the classical theory of particles 
and fields} (Springer, Heidelberg, 2007).

\bibitem{H1}
P. Havas, Phys. Rev. {\bf 87}, 309 (1952).

\bibitem{BV}
A. O. Barut and D. Villarroel, J. Phys. A: Math. Gen. {\bf 8}, 156 (1975). 

\bibitem{Bha}
H. J. Bhabha, Proc. R. Soc. (London) {\bf A172}, 384 (1939).

\bibitem{CM}
R. G. Cawley and E. Marx, Int. J. Theor. Phys. {\bf 1}, 153 (1968).

\bibitem{Rohr}
Rohrlich F., {\it Classical Charged Particles} (Addison-Wesley, 
Redwood, CA, 1990). 

\bibitem{Cw69}
R. G. Cawley, Ann. Phys., NY {\bf 54}, 122 (1969).

\bibitem{Cw70}
R. G. Cawley, J. Math. Phys. {\bf 11}, 761 (1970).

\bibitem{LV}
C. A. L\'opez and D. Villarroel, Phys. Rev. D {\bf 11}, 2724 (1975).

\bibitem{PoisPr}
E. Poisson, An introduction to the Lorentz-Dirac equation, 
arXiv:gr-qc/9912045 (1999). 

\bibitem{Y4D}
Yu. Yaremko, J.Phys.A: Math.Gen. {\bf 36}, 5149 (2003).

\bibitem{Y6D}
Yu. Yaremko, J.Phys.A: Math.Gen. {\bf 37}, 1079 (2004).

\bibitem{HvC}
F. R. Jr. Crownfield and P. Havas, Phys. Rev. {\bf 94}, 471 (1954).

\bibitem{BHC44}
H.J. Bhabha and Harish-Chandra, Proc. R. Soc. (London) {\bf A183}, 134
(1944).

\bibitem{BHC}
H. J. Bhabha and Harish-Chandra, Proc. R. Soc. (London) {\bf A185}, 250 
(1946). 

\bibitem{BHPs}
L. M. Burko, A. I. Harte, and E. Poisson, Phys. Rev. D {\bf 65}, 124006
(2002).

\bibitem{HPs}
R. Haas and E. Poisson, Class. Quantum Grav. {\bf 22}, S739 (2005).

\bibitem{B}
L. M. Burko, Class. Quantum Grav. {\bf 19}, 3745 (2002).

\end{thebibliography}
\end{document}